# Towards a Rosetta Stone for (meta)data: Learning from natural language to improve semantic and cognitive interoperability

Vogt, Lars[1] orcid.org/0000-0002-8280-0487;

Konrad, Marcel[1] orcid.org/0000-0002-2452-3143

Prinz, Manuel[1] orcid.org/0000-0003-2151-4556

[1] *TIB Leibniz Information Centre for Science and Technology, Welfengarten 1B, 30167 Hanover, Germany*

Corresponding Author: lars.m.vogt@gmail.com

# Abstract

In order to effectively manage the overwhelming influx of data, it is crucial to ensure that data is findable, accessible, interoperable, and reusable (FAIR). While ontologies and knowledge graphs have been employed to enhance FAIRness, challenges remain regarding semantic and cognitive interoperability. We explore how English facilitates reliable communication of terms and statements, and transfer our findings to a framework of ontologies and knowledge graphs, while treating terms and statements as minimal information units. We categorize statement types based on their predicates, recognizing the limitations of modeling non-binary predicates with multiple triples, which negatively impacts interoperability. Terms are associated with different frames of reference, and different operations require different schemata. Term mappings and schema crosswalks are therefore vital for semantic interoperability. We propose a machine-actionable Rosetta Stone Framework for (meta)data, which uses reference terms and schemata as an *interlingua* to minimize mappings and crosswalks. Modeling statements rather than a human-independent reality ensures cognitive familiarity and thus better interoperability of data structures. We extend this Rosetta modeling paradigm to reference schemata, resulting in simple schemata with a consistent structure across statement types, empowering domain experts to create their own schemata using the Rosetta Editor, without requiring knowledge of semantics. The Editor also allows specifying textual and graphical display templates for each schema, delivering human-readable data representations alongside machine-actionable data structures. The Rosetta Query Builder derives queries based on completed input forms and the information from corresponding reference schemata. This work sets the conceptual ground for the Rosetta Stone Framework that we plan to develop in the future.

# Introduction

In the current era, we are witnessing an exponential growth in the generation and consumption of data. Statista's projections from 2020 indicate that the global data creation, capture, copying, and consumption will reach an estimated 97 zettabytes in 2022, with a staggering daily volume of approximately 328.77 billion terabytes by 2023, reflecting a notable 23.71% increase over the preceding year, with videos accounting for over 50% of the global traffic (1,2). Additionally, data production in the last two years accounted for a striking 90% of the entire data generated worldwide in 2016 (3), illustrating the rapid pace of data expansion. The doubling of the overall data volume every three years (4) further accentuates the magnitude of this exponential growth. Simultaneously, the scholarly domain is witnessing a surge in publications, with an annual output of over 7 million academic papers (5). These figures emphasize the **urgency of harnessing machine support, as without it, the sheer volume of (meta)data threatens to overwhelm and impede meaningful insights**.

With this in mind, it is essential to facilitate **machine-actionable (meta)data** in scientific research, so that machines assist researchers in identifying relevant (meta)data pertaining to a specific research question. Moreover, enhancing the machine-actionability of (meta)data offers a potential solution to the reproducibility crisis in science (6), enabling the availability, findability, and usability of raw (meta)data (7).

Considering the substantial global investment in research and development, which reached a staggering sum of $1,609,214 billion in 2021 (8), failure to render the corresponding output of (meta)data machine-actionable could potentially result in redundant research efforts. By enabling machine-actionable (meta)data, substantial savings can be achieved, redirecting resources towards novel and non-repetitive research endeavors. However, the concept of machine-actionable (meta)data requires clarification.

---

**Box 1 | Machine-Readability, Machine-Interpretability, Machine-Actionability** (9,10)

**Machine-readable** are those elements in bit-sequences that are clearly defined by structural specifications, such as data formats like CSV, JSON, or XML, or resources and literals in the Resource Description Framework (RDF).

**Machine-interpretable** are those elements in bit-sequences that are **machine-readable** and can be related with semantic artefacts in a given context and therefore have a defined purpose, such as referencing defined and registered ontology terms that provide meaning to a resource in an RDF triple following the triple syntax of *Subject-Predicate-Object*.

**Machine-actionable** are those elements in bit-sequences that are **machine-interpretable** and belong to a type of element for which operations have been specified in symbolic grammar, thus linking types of (meta)data statements to operations such as logical reasoning based on description logics for OWL-based (meta)data and other rule-based operations such as unit conversion or other data conversions.

---

Looking at the definition of **machine-actionable** (see Box 1), it is evident that this attribute **cannot be simplified as a mere Boolean property**. Instead, it exists on a spectrum, allowing for **degrees of machine-actionability**. Numerous operations can potentially be applied to a given set of (meta)data, and the ability to apply even a single operation would suffice to classify the (meta)data as machine-actionable. Consequently, specifying the set of operations that can be applied to the (meta)data, along with the corresponding tools or code, is more meaningful than labeling the entire set as machine-actionable. For instance, one could state that *"Dataset A is machine-actionable with respect to operation X using tool Y"*.

It is worth noting that reading a dataset could be considered an operation itself. Therefore, datasets documented in formats as PDF, XML, or even ASCII files can be considered machine-readable and, to some extent, already machine-actionable. Moreover, if a dataset is machine-readable, search operations can be performed on it, enabling the identification of specific elements through label matching, for example. The success of search operations serves as a measure of the findability of (meta)data. Similarly to machine-actionability, **findability cannot be characterized as a Boolean property**. Machine-readable (meta)data can be found through label matching, while interpretable (meta)data can be found through their meaning, referent, or contextual information. Thus, (meta)data that are readable but not interpretable possess limited findability. It is important to emphasize that our definition of machine-actionability, as outlined in Box 1, strictly depends on machine-interoperability. Consequently, machine-reading of a dataset and machine-searching based on label matching are not considered proper examples for operations that fulfill the requirements of machine-actionability.

In 2016, the **FAIR Guiding Principles for (meta)data** were introduced, providing a framework to assess the extent to which (meta)data are F̲indable, A̲ccessible, I̲nteroperable, and R̲eusable for both machines and humans alike (11). These principles have gained increasing attention from the research, industry, and knowledge management tool development communities in recent years (11–16). Furthermore, stakeholders in science and research policy have recognized the significance of the FAIR Principles. The economic impact of FAIR research (meta)data was estimated by the European Union (EU) in 2018, revealing that the lack of FAIR (meta)data costs the EU economy at least 10.2 billion Euros annually. Taking into account the positive effects of FAIR (meta)data on data quality and machine-readability, an additional 16 billion Euros were estimated (17). Consequently, the High Level Expert Group of the [European Open Science Cloud](#) (EOSC) recommended the establishment of an Internet of FAIR Data and Services ([IFDS](#)) (18). The IFDS aims to enable data-rich institutions, research projects, and citizen-science initiatives to make their (meta)data accessible in accordance with the FAIR Guiding Principles, while retaining control over ethical, privacy, and legal aspects of their (meta)data (following Barend Mons' *data visiting as opposed to data sharing* (19)). Achieving this goal requires the provision of rich machine-actionable (meta)data, their organization into **FAIR Digital Objects** (20,21), each identifiable by a Unique Persistent and Resolvable Identifier (UPRI), and the development of suitable concepts and tools for human-readable interface outputs and search capabilities. Although progress has been made toward building the IFDS (see the [GO FAIR Initiative](#) and [EOSC](#)), the current state of the FAIRness of (meta)data in many data-rich institutions and companies is still far from ideal.

The increasing volume, velocity, variety, and complexity of data present significant challenges that traditional methods and techniques for handling, processing, analyzing, managing, storing, and retrieving (meta)data struggle to address effectively within a reasonable timeframe (22). However, **knowledge graphs** in conjunction with **ontologies**, offer a promising technical solution for implementing the FAIR Guiding Principles, thanks to their transparent semantics, highly structured syntax, and standardized formats (23,24). Knowledge graphs represent instances, classes, and their relationships as resources with their own UPRIs. These UPRIs are employed to denote relationships between entities using the triple syntax of *Subject-Predicate-Object*. Each particular relationship is thus modeled as a structured set of three distinct data points. In contrast, relational databases model entity relationships between data table columns and not between individual data points. Consequently, knowledge graphs outperform relational databases in handling complex queries on densely connected data, which is often the case with research (meta)data. Therefore, knowledge graphs are particularly well suited for FAIR research and development, as well as any task requiring efficient and detailed retrieval of (meta)data.

Nonetheless, employing a knowledge graph to document (meta)data does not guarantee adherence to FAIR Principles. Achieving FAIRness necessitates additional guidelines, such as consistent usage of the same semantic data model for identical types of (meta)data statements to ensure schematic interoperability (10), as well as organizing (meta)data into FAIR Digital Objects (20,21). Knowledge graphs, being a relatively new concept and technology, introduce their own specific technical, conceptual, and societal challenges. This is evident in the somewhat ambiguous nature of the knowledge graph concept (23) and the absence of commonly accepted standards, given the diverse technical and conceptual incarnations ranging from labeled property graphs like [Neo4J](#) to approaches based on the Resource Description Framework (RDF), employing RDF stores and applications of description logic using the Web Ontology Language (OWL).

Regardless of these considerations, it can be argued that the demand for FAIR and machine-actionable (meta)data presupposes their **successful communication between machines and between humans and machines**. During such communication processes, preserving the meaning and reference of the message between sender and receiver is crucial, requiring both parties to share the same background knowledge, encompassing lexical competences, syntax and grammar rules, and relevant contextual knowledge.

In this paper, we lay the conceptual groundwork for a future machine-actionable Rosetta Stone Framework designed to achieve cognitive and semantic interoperability of (meta)data. We begin by examining the notion of interoperability and emphasizing the necessity of cognitive interoperability. Drawing inspiration from natural languages like English, we explore how semantic interoperability can be understood by analyzing the way terms and statements convey meaning and information. Recognizing terms and statements as basic units of information, we investigate the linguistic structures that ensure reliable information communication and draw parallels to the structures found in (meta)data schemata. This analysis yields insights into achieving semantic interoperability of terms and statements.

Our main argument centers around the requirement for a machine-actionable Rosetta Stone Framework for (meta)data, which addresses terminological and propositional interoperability. This framework leverages reference terms and reference schemata as an *interlingua* to establish term mappings and schema crosswalks, facilitating cognitive and semantic interoperability. The emphasis lies on a modeling paradigm that enables machine-interpretability of (meta)data, prioritizing their findability and interoperability over reasoning capabilities. This prioritization opens up new avenues for modeling by shifting away from the paradigm frequently applied in science that focuses on modeling a human-independent reality. Instead, the Rosetta Modeling Paradigm models the structure of statements to enhance efficient and reliable information communication.

We propose two alternative versions of a Rosetta Modeling Approach: a minimal light version and a full version that supports versioning of statements and tracks the editing history of each statement within a knowledge graph. Furthermore, we discuss the specifications for a Rosetta Editor, a tool that will allow domain experts to create new reference schemata without requiring expertise in semantics or computer science. The Rosetta Editor will also enable the specification of display templates for human-readable text-based and graph-based representation of (meta)data.

Additionally, we propose a Rosetta Query Builder that allows users to write queries without the need for a graph query language. Finally, we provide an overview of the Rosetta Framework and explore related work in the field before presenting our concluding remarks.

It is important to note that this paper represents conceptual work and serves as a foundation for future development, testing, and implementation of the envisioned Rosetta Stone Framework.

> **Box 2 | Conventions**
>
> In this paper, we refer to FAIR knowledge graphs as machine-actionable semantic graphs for documenting, organizing, and representing lexical, assertional (e.g., empirical data), universal, and contingent statements and thus a mixture of ABox and TBox expressions (thereby contrasting knowledge graphs with ontologies, with the latter containing mainly universal statements and thus TBox expressions and lexical statements). We use *triple* to refer to triple statements, and *statement* to refer to natural language statements.
>
> For reasons of clarity, in the text and in all figures we represent resources not with their UPRIs but with human-readable labels, with the implicit assumption that every property, every instance, and every class has its own UPRI.

# Interoperability

Interoperability is directly dependent on machine-actionability: datasets *A* and *B* can be said to be interoperable if there is an operation *X* that can be applied equally to both. Because of this dependency, **interoperability inherits from machine-actionability that it is not a Boolean property, but rather describes a continuum**, and its degree depends on the number of operations that can be applied to a given type of (meta)data. Interoperability also involves the ability to identify the type of (meta)data within a given dataset that can be processed by a given operation, and vice versa.

(Meta)data are composed of terms (here, meant in a broad sense, including also symbols and values) that form statements. Both, terms and statements, carry meaning, and thus semantic content, and both are required for the successful communication of information. The interoperability of terms and statements between a sender and a receiver of (meta)data is therefore a prerequisite for their successful communication. Successfully communicating (meta)data between machines and between a machine and a human being requires not only their successful transmission, so that the receiver can read them (i.e., readability), but also their successful processing, so that the receiver understands their meaning (i.e., interpretability) and can use them in another context by applying specific operations to them (i.e., actionability).

Obviously, interoperability plays a very important role in this communication process and is also central to the realization of FAIR (meta)data. Without interoperability, the findability and reusability of (meta)data is limited, and without interoperability there is no machine-actionability. This central role of interoperability has also been recognized by the EOSC. In their **EOSC Interoperability Framework** (20), they distinguish four layers of interoperability:

- **technical interoperability** (i.e., information technology systems must work with other information technology systems in implementation or access without any restrictions or with controlled access),
- **semantic interoperability** (i.e., contextual semantics related to common semantic resources),
- **organizational interoperability** (i.e., contextual processes related to common process resources), and

- **legal interoperability** (i.e., contextual licenses related to common license resources).

However, interoperability is not only required for the successful communication of information between machines. It is also required for the communication of information between humans and machines, and thus involves an additional layer of interoperability that takes into account the need for communication to follow and apply rules, and that requires knowledge to be shared between humans and machines, involving the correct use of devices and tools that humans need to learn and understand. We believe that another layer of interoperability, i.e., cognitive interoperability, needs to be added to the EOSC Interoperability Framework. We understand **cognitive interoperability** as characterized in Box 3.

> **Box 3 | Cognitive Interoperability**
>
> Cognitive interoperability is "a characteristic of an information technology system to communicate its information with human users in the most efficient ways, providing them with tools and functions that intuitively support them in getting an overview of the data, find the data they are interested in, and allow them to explore other data points from a given data point in semantically meaningful and intuitive ways, thereby taking into account the general cognitive conditions of human beings—not only in terms of how humans prefer to interact with technology (human-computer interaction) but also in terms of how they interact with information (human-information interaction). In the context of [knowledge graphs], the tools should make the user aware of their contents, help to understand their meaning, support communicating their contents, provide means that increase the trustworthiness of their contents, support their integration in other workflows and software tools, and make transparent what actions can be taken. Additionally, cognitive interoperability of an information technology system can also be understood as the system's characteristic to be easily implemented and employed by developers and easily managed by operators" (p. 8-9) (10).

Cognitive interoperability focuses on the usability of data structures and knowledge management systems for human users and developers—an aspect that has been somewhat overlooked, especially in the context of knowledge graphs and semantic technologies. Cognitive interoperability also means taking into account how humans typically communicate. As experts in communicating information efficiently, omitting background knowledge and using somewhat fuzzy statements that refer to general figures of thought and that use metaphors and metonymies, we humans are usually very efficient in our communication, reducing the information to be communicated to a minimum, knowing that other human beings will still be able to understand and infer the missing information from the context. For a machine, on the other hand, all relevant information must be explicitly stated. As a consequence, cognitive interoperability has to deal with the dilemma that arises from the conflict between machine-actionability and human-actionability of (meta)data representations: **the more data representations are pushed towards machine-actionability, the more complex they become and thus the less human-actionable** (10) (see Fig. 1, middle). This is an impedance mismatch. It has the potential to frustrate humans communicating (meta)data with machines.

**Observation:**

*This apple has a weight of 212.45 grams, with a 95% conf. int. of 212.44 to 212.47 grams.*

**Machine-Actionable RDF Graph:**

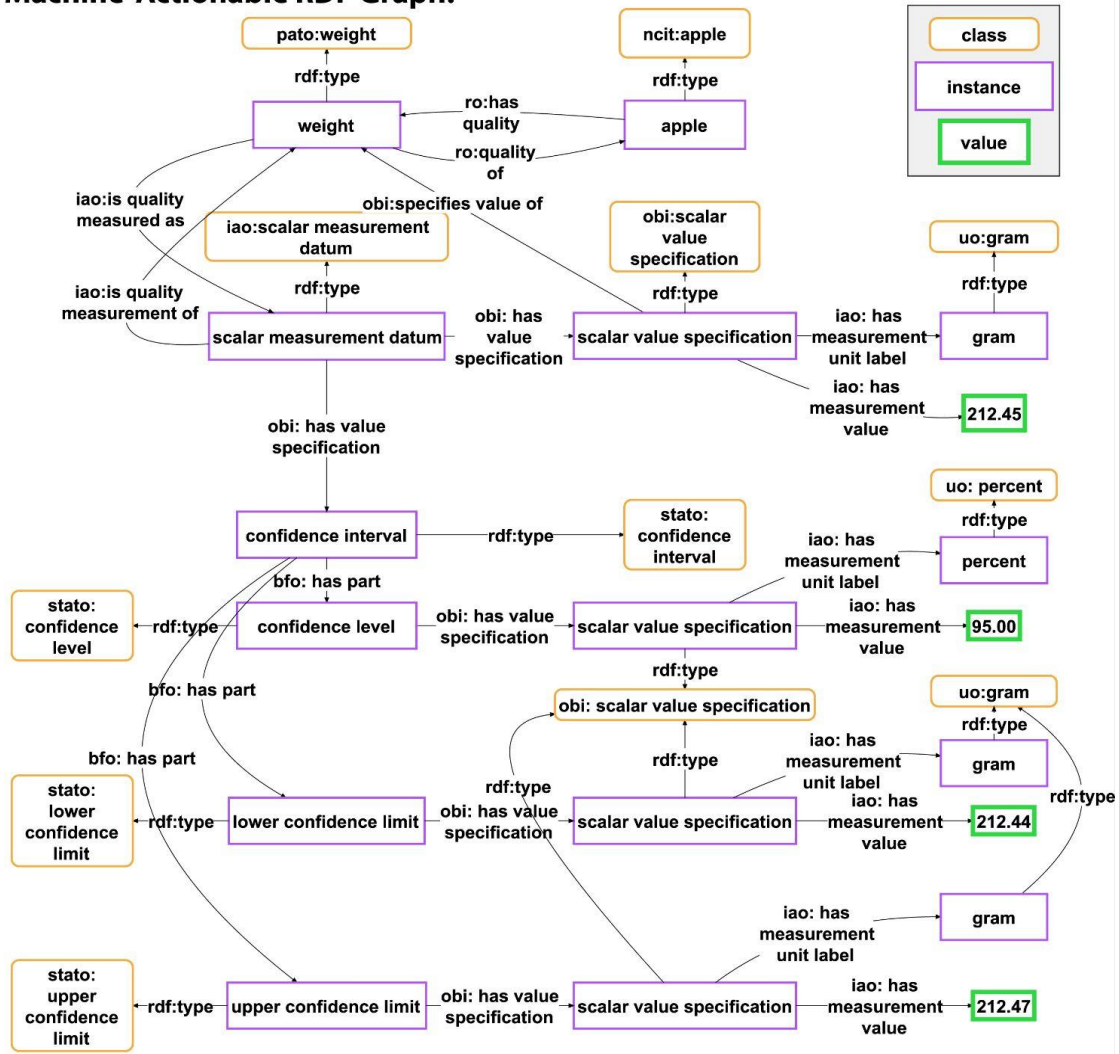

**Human-Actionable Mind-Map like Graph:**

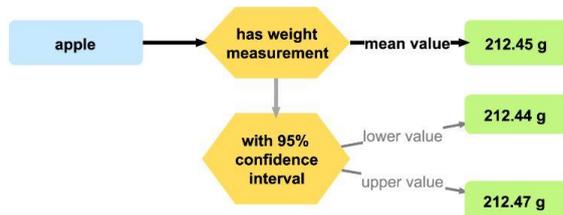

**Figure 1: Comparison of a human-readable statement with its machine-actionable representation and its human-actionable representation. Top**: A human-readable statement about the observation that a particular apple weighs 212.45 grams, with a 95% confidence interval of 212.44 to 212.47 grams. **Middle**: A machine-actionable representation of the same statement as an ABox semantic graph, using RDF syntax and following the general pattern for measurement data from the Ontology for Biomedical Investigations (OBI) (25) of the Open Biological and Biomedical Ontology (OBO) Foundry. **Bottom**: A human-actionable representation of the same statement as a mind-map like graph, reducing the complexity of the RDF graph to the information that is actually relevant to a human reader. [Figure taken from (10)]

If we want to store (meta)data in a knowledge graph in a machine-actionable format and at the same time present them in an easily understandable, human-readable way in the user interface (UI), we need to **decouple the data storage in the graph from the data presentation in the UI**, so that information that is only necessary for machines but irrelevant for humans is only accessed by machines but not displayed in the UI (see Fig. 1, bottom). However, considering the complexity of the tasks that users of a knowledge graph want to accomplish (e.g., fact-finding, understanding cause-effect chains, or understanding controversial topics) (26), it becomes clear that cognitive interoperability of (meta)data involves more than this. The cognitive interoperability of a graph can also be enhanced by developing new approaches and tools for **exploring and navigating the graph**, zooming in and out at different levels of representational granularity, thereby reducing the complexity of the graph to only those bits of information that are currently relevant to the user. This adds another layer to the requirements placed on the UIs of FAIR knowledge graphs to support different user tasks, following **visual information seeking mantras** such as '*overview first, zoom and filter, then details-on-demand*' (27), '*search first, show context, and expand on demand*' (28), and '*details first, show context, and overview last*' (29) or '*overview for navigation*' (10).

Tools for describing graph patterns that enforce a standardized way of modeling and representing data of the same type, such as the Shapes Constraint Language [SHACL](#) and Data Shapes [DASH](#) (30), Shape Expressions [ShEx](#) (31,32), or the Reasonable Ontology Templates [OTTR](#) (33,34), provide some support for decoupling data storage from data presentation. However, they do not support developers and data stewards in using (meta)data or writing queries against them, nor do they provide the structure to the graph required to support semantically meaningful navigation and exploration of the graph. Unfortunately, the FAIR Guiding Principles do not address the need to decouple data storage from data presentation, nor do they address the need to explore (meta)data in a human-actionable way. Therefore, we proposed to extend the FAIR Guiding Principles to include the **principle of human explorability**, resulting in the **FAIREr Guiding Principles** (10), which take into account the cognitive interoperability of (meta)data (for a detailed discussion see (10)).

# Semantic interoperability and what natural languages like English can teach us

The EOSC Interoperability Framework (20) characterizes semantic interoperability as a requirement for enabling machine-actionability between information systems, and it is achieved *"when the information transferred has, in its communicated form, all of the meaning required for the receiving system to interpret it correctly"* (p. 11).

To understand what semantic interoperability means at a conceptual level, it is helpful to consider how we as humans communicate meaning (i.e., semantic content) in a natural language such as English, using terms and statements as the basic units of meaning that carrying information. And when we talk about communication, we mean the attempt to create the same cognitive representation of information in the receiver as is present in the sender.

## Requirements for successfully communicating terms and statements

What is needed to communicate terms and statements efficiently and reliably? For communication between human beings to be successful, both the sender and the receiver of the information need to share the same relevant background knowledge. For a given **term**, sender and receiver have to share the same lexical competences. They need to share the same **inferential lexical competences** (35) and thus knowledge about the **meaning** of a given term in the form of an **ontological definition** that answers the question '*What is it?*' (36). Inferential lexical competences are needed for the **human-interpretability of terms**.

Sender and receiver also need to share the same **referential lexical competences** (35) and thus **diagnostic knowledge** about the **reference** of a given term. A **proper name** refers to an individual entity (i.e., a particular) and a **general term** or **kind term** to a set of individuals that meet the defining properties of the term, sometimes also called its extension. A **verb** or **predicate** refers to a specific type of action or attribute. Diagnostic knowledge is often communicated in the form of method-dependent recognition criteria, images, or exemplars that answer the question '*How does it look, how to recognize or identify it?*' (36), enabling the receiver to use the term correctly in **designation** (i.e., the object is given, and the matching term has to be found) or **recognition tasks** (i.e., the term is given, and the matching object has to be identified)[1]. Referential lexical competences are needed to use a term correctly in different contexts, and thus for the **human-actionability of terms**.

Given that the meaning of terms is provided by their ontological definitions, one could argue that terms are only placeholders, i.e., surrogates, for these definitions and thus statements, and that only statements carry meaning. In other words, the communication of meaning requires more than a single term, but rather several terms placed in context and related by predicates. The communication of meaning thus requires statements. **Statements carry meaning in addition to the meaning of the terms that compose them.** This becomes obvious when the positions of terms in a given sentence are changed, such as *"Peter travels from Berlin to Paris"* versus *"Peter travels from Paris to Berlin"*—the same set of terms carries two different meanings. Therefore, for the efficient and reliable communication of statements, the sender and the receiver of the information must share a set of rules and conventions for formulating sentences using terms.

But how is the meaning of a statement represented in the human brain? Understanding how the human brain creates cognitive representations of semantic content is a very active area of research (38). The human brain is a highly interconnected complex system that is continually influenced by input signals from the body and the world, so that a given neuron does not function in isolation but is substantially influenced by its neural context (39). It is therefore not surprising that there is evidence for at least two forms of object knowledge representation in the human brain, supported by different brain systems, i.e., motor-sensory-derived and conceptual-cognition-derived knowledge (40,41), and that lexical concepts are stored as patterns of neural activity that are encoded in multiple areas of the brain, including taxonomic and distributional structures as well as experience-based representational structures that encode information about sensory-motor, affective, and other features of phenomenal experience (42). These findings suggest that the cognitive representation of the meaning of a statement is likely to take the form of a complex network of associations, analogous to a multidimensional mind-map. Thus, when attempting to communicate a statement, the sender must first translate this multidimensional mind-map into a one-dimensional sequence of terms, i.e., a sentence. This translation step is supported by a set of syntactic and grammatical conventions, shared by the sender and the receiver, for formulating sentences using terms.

---

[1] Unfortunately, many ontologies only provide ontological definitions, not recognition criteria. For example, the term '*cell nucleus*' (FMA:63840) of the Foundational Model of Anatomy ontology (37) is defined as "*Organelle which has as its direct parts a nuclear membrane and nuclear matrix*". As we cannot see cell nuclei without using a microscope and staining a cell sample, the definition does not provide the practical diagnostic knowledge needed for designation and recognition tasks. The term therefore does not provide the information needed to build the referential lexical competences required to make the term human-actionable.

According to the **predicate-argument-structure** of linguistics (43,44), the main verb of a statement, together with its auxiliaries, forms the statement's predicate. A predicate has a **valence** that determines the number and types of **arguments** it requires to complete its meaning. **Adjuncts** may be additionally related to the predicate, but they are not necessary to complete the predicate's meaning. Adjuncts provide optional information, such as a time specification in a parthood statement. Therefore, every statement has a subject phrase as one of its arguments, and can have one or more object phrases as further arguments and additional adjuncts, depending on the underlying predicate.

In the syntax of a statement, each argument and adjunct of a predicate can be understood as having a specific **position** with a specific **semantic role**. This is related to Kipper et al.'s (45) verb lexicon [VerbNet](), where they extend Levin verb classes (46) to include abstract representations of syntactic frames for each class, with explicit correspondences between **syntactic positions** (i.e., positions in a syntax tree) and the **semantic roles** (i.e., thematic roles sensu (45)) that these positions express (Fig. 2B). Each verb class lists the semantic roles allowed by the predicate-argument-structure of the instances of the class and the basic syntactic frames shared by the instances.

The list of arguments of a predicate-argument-structure can be described by a list of **thematic labels** taken from a set of predefined possible labels (e.g., AGENT, PATIENT, THEME, etc.), and the syntactic frames are represented by an ordered sequence of such thematic labels. The thematic labels function as descriptors of semantic roles that are mapped onto positions in a given syntactic frame (45).

In [PropBank](), Palmer et al. (47) define semantic roles at the level of individual verbs by numbering the arguments of the verb, with the first argument generally taking the semantic role of an AGENT and the second argument typically, but not necessarily, taking the semantic role of a PATIENT or a THEME. For higher numbered arguments, there are no consistently defined roles.

Syntax trees, with their different syntactic positions and associated semantic roles, contribute substantially to the meaning of their sentences, and they are used to translate a web of ideas in the mind of a sender into a string of words that can be understood by a receiver to translate it back into a web of ideas (48). The clearer the semantic roles of the different positions are, the easier it is for a human being to understand the information. In a sense, we can understand syntax trees as the first knowledge graphs created by humans, and their use seems to be quite straightforward, providing a structure that is interoperable with human cognitive conditions, thus satisfying the need for **cognitive interoperability**.

To sum it up, whenever information needs to be communicated efficiently and reliably, the sender and receiver of the information must not only share the same inferential and referential lexical competences regarding the terms used in their communication, but also the same set of syntactic and grammatical conventions for formulating sentences with these terms, resulting in the same syntax tree in both sender and receiver.

### A) Natural Language Statement

*This apple has a weight of 212.45 grams.*

### B) Syntax Tree with Syntactic Positions and associated Semantic Roles

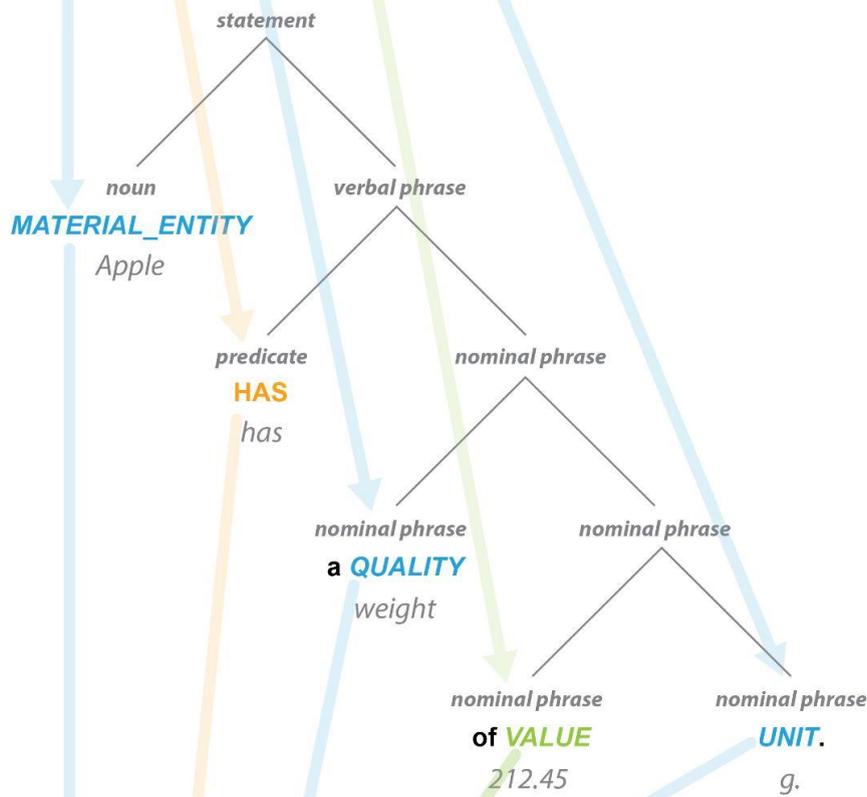

### C) Formalized Statement with Syntactic Positions and associated Semantic Roles

**OBJECT HAS** a **QUALITY** of **VALUE UNIT**.

### D) Tabular Data Schema with Slots and Constraints

| OBJECT | QUALITY | VALUE | UNIT |
|---|---|---|---|
| apple | weight | 212.45 | gram |

### E) Graph Data Schema with Slots and Constraints

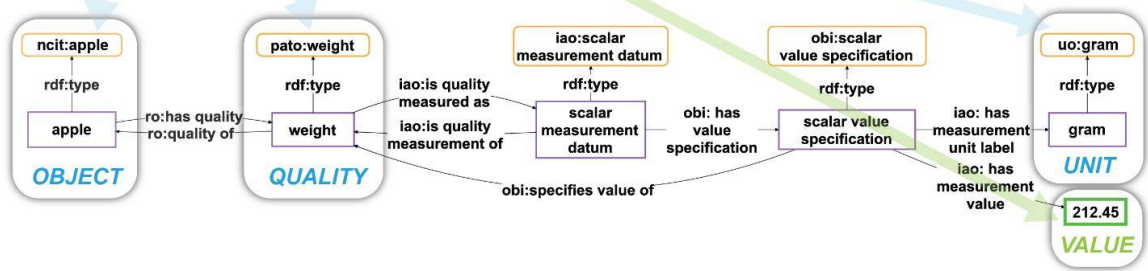

**Figure 2: Parallels between natural language statements and data schemata.** The natural language statement in **A)** is structured by syntactical and grammatical conventions into syntactic positions of phrases of a syntax tree as shown in **B)** or a formalized statement as is shown in **C)**, where each position having a specific semantic role associated with it that can be described by a thematic label. Data schemata, both tabular as in **D)** and graphical as in **E)**, must represent these syntactic positions in the form of slots, and each slot must specify its associated semantic role in the form of a constraint specification.

The knowledge shared by the sender and the receiver of the message ensures that the message is:

1. **readable**: The receiver must be able to **identify the basic units of information** in a message, i.e., where a term or a statement begins and where it ends;
2. **interpretable**: The receiver must be able to understand each term and statement and thus recognize its **meaning**;
3. **actionable**: The receiver must be able to correctly apply terms and statements from the message, including correctly designating or recognizing **referents** of terms and correctly placing a statement in **context**.

# Parallels between the structure of natural language statements and data schemata with implications for semantic interoperability

Since we can think of each datum as a somewhat formalized representation of a natural language statement, structured in such a way that it can be easily compared with statements of the same type, and easily read and operationalized by machines (cf. Fig. 2A with 2D,E), we can think of a data schema as a formalization of a particular type of natural language statement to support its machine-actionability. In other words, data schemata are to machines what syntax trees are to humans—both define positions with semantic roles. When we compare data schemata with their corresponding natural language statements, we can thus see similarities between the structure of a sentence defined by the syntax and grammar of a natural language and the structure of a corresponding schema (Fig. 2). As discussed above, the **syntactic positions** of terms in a natural language sentence take on specific **semantic roles** and contribute significantly to the meaning of the statement. For a data schema to have the same meaning as its corresponding natural language statements, it must, as a **minimum requirement**, functionally and semantically provide a similar structure with the same elements as the corresponding syntax tree: the schema must represent all relevant syntactic positions—in schemas often called **slots**—and their associated semantic roles in the form of **constraint specifications**, with terms and values populating the slots (see Fig. 2D,E). After all, humans need to be able to understand these data schemata, and need to be able to translate a given datum represented in a given data schema back into a natural language statement. Data schemata can therefore be seen as attempts to translate the structure of natural language sentences into machine-actionable data structures.

Looking at these interdependencies, we can distinguish different **causes for the lack of semantic interoperability**. And since we distinguish between terms and propositions as two different types of meaning-carrying entities when we communicate (meta)data, we can distinguish causes for the lack of terminological interoperability from causes for the lack of propositional interoperability.

With regard to **terminological interoperability**, we can distinguish between ontological and referential causes for the lack of interoperability of terms (10). When two given terms are compared semantically, they can either (i) differ in both their meaning and their referent (e.g., '*apple*' and '*car*'), (ii) differ only in their meaning but share the same referent (e.g., '*Morning Star*' and '*Evening Star*', both referring to the planet Venus), or (iii) share the same meaning and the same referent.

If two terms share their meaning and their referent, they are **referentially and ontologically interoperable**. They are strict synonyms and can be used interchangeably. Since no two terms can share their meaning but not their referent, ontological interoperability always implies referential interoperability, but not vice versa. Thus, if two terms have the same referent but not the same meaning, because controlled vocabularies may differ in their ontological commitments, their ontological interoperability is violated, but not necessarily their **referential interoperability**, since both terms can be used to refer to the same entity. For example, the COVID-19 Vocabulary Ontology (*COVoC*) defines 'viruses' as a subclass of 'organism', while the Virus Infectious Diseases Ontology (*vido*) defines 'virus' as a subclass of 'acellular structure', and thus as an object that is not an organism–these two terms are therefore not ontologically interoperable, even though they have the same referent (i.e., the same extension). In other words, the set of ontologically interoperable terms is a subset of the set of referentially interoperable terms.

As far as terminological interoperability is concerned, we can therefore conclude that although ontological interoperability is preferred[2], **referential interoperability is the minimum requirement for the interoperability of terms**, since when we communicate information, we at least want to know that we are referring to the same entities.

In the context of knowledge graphs and ontologies, if two terms share the same meaning and referent despite having different UPRIs, we can express their terminological interoperability by specifying a corresponding term mapping using the '*same as*' (owl:sameAs) property. If two terms have only the same referent but not the same meaning, we can express their referential interoperability by specifying a corresponding term mapping using the '*equivalent class*' (owl:equivalentClass) property. **We can thus distinguish between ontological (i.e., same-as) and referential (i.e., equivalent-class) term mappings**. Both types of mappings are **homogeneous definition mappings** (49), where there is only one vocabulary element to be mapped on the left side, and several others on the right side of the definition that do not need to be mapped.

With respect to **propositional interoperability**, we can distinguish between logical and schematic causes for the lack of interoperability of statements of the same type (10). Data and metadata statements are **logically interoperable** if they have been modeled on the basis of the same logical framework (e.g., OWL-Full and thus on full description logics), so that one can reason over them using appropriate reasoners (e.g., an OWL-Full reasoner). When we talk about logically interoperable terms, we are actually referring to the ontological definitions of these terms and thus to their class axioms, which are universal statements that can be placed in the same logical framework if the terms are logically interoperable.

**Schematic interoperability** is achieved when statements of the same type are documented using the same (meta)data schema. If statements of the same type were represented using different schemata, corresponding (meta)data would no longer be interoperable. In such cases, one would have to specify schema crosswalks by aligning slots (i.e., syntactic positions) that share the same constraint specification (i.e., semantic roles) across different schemata modeling the same type of statement, in order to regain schematic interoperability (see Fig. 3). If the schemata use different vocabularies to populate their slots (i.e., the constraint specifications refer to different vocabularies), then corresponding term mappings must be included in the crosswalk to ensure terminological interoperability (see red bordered slots in Fig. 3). Consequently, we can distinguish between **ontological and referential schema crosswalks**. A schema crosswalk is a set of rules that specifies how (meta)data elements or attributes (i.e., slots) from one schema and format can be aligned and mapped to the equivalent (meta)data elements or attributes in another schema and format.

---

[2] In order to achieve ontological interoperability, however, the schemata used for the statements in the class axioms of the terms to be mapped (i.e., their ontological definitions) must also be aligned if they differ. Furthermore, if the terms mentioned in these class axioms differ between the terms to be mapped, they also need to be mapped if they have the same meaning and referent(s). This is necessary, though often overlooked, because the meaning of a term is communicated by its definition, which is a statement in its own right, with all the consequences for semantic interoperability that this implies.

In other words, to achieve schematic interoperability between two given statements, the subject, predicate, and object slots of their (meta)data schemata need to be aligned, and their terms be mapped across controlled vocabularies. To do this, the schemata must first be formally specified, e.g., in the form of graph patterns specified as SHACL shapes. Shapes that share the same statement-type referent then need to be aligned and mapped. These are **ontology pattern alignments** (i.e., TBox alignments) (49) or **ABox alignments**, where several vocabulary elements must first be aligned and then mapped in schema crosswalks.

As far as propositional interoperability is concerned, we can therefore conclude that although a combination of logical interoperability and schematic interoperability is preferred, for the same reasons as for terms, **schematic interoperability using referential schema crosswalks is the minimum requirement for the interoperability of statements**.

In summary, the interoperability of (meta)data statements does not only depend on the number of applicable operations and thus on machine-actionability, but also on the completeness of the ontological and referential term mappings and schema crosswalks relevant to the statements.

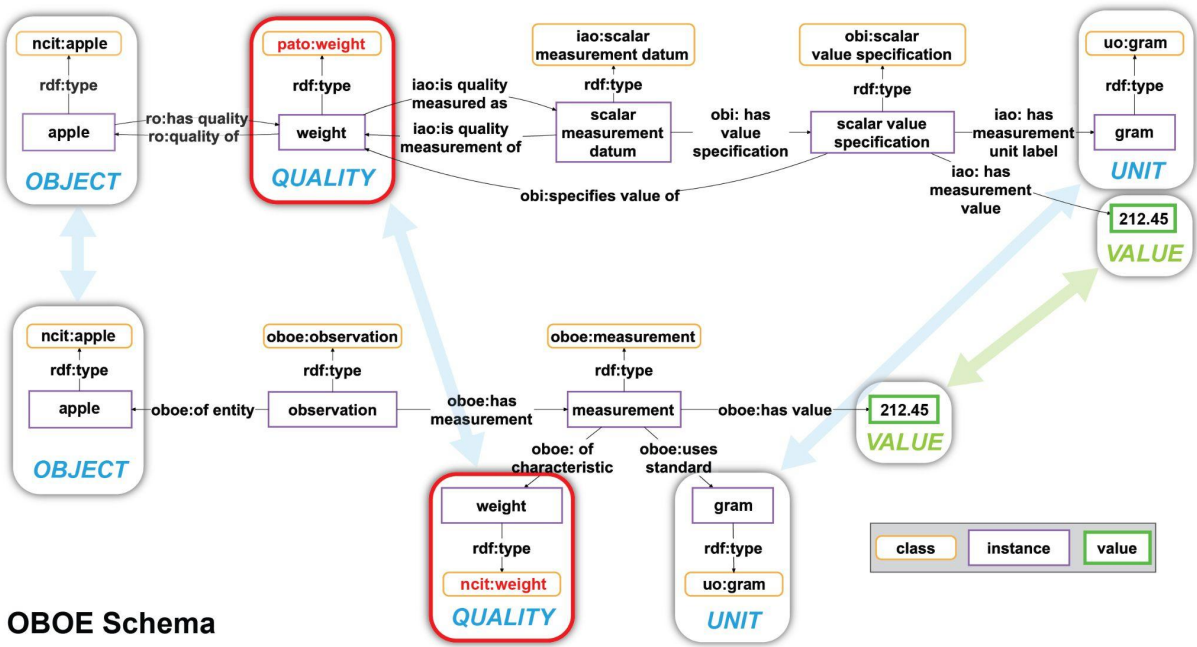

**Figure 3: Crosswalk from one schema to another for a weight measurement statement.** The same weight measurement statement is modeled using two different schemata. **Top**: The weight measurement according to the schema of the Ontology for Biomedical Investigations (OBI) (25) of the Open Biological and Biomedical Ontology (OBO) Foundry, which is often used in the biomedical domain. **Bottom**: The same weight measurement according to the schema of the Extensible Observation Ontology (OBOE), which is often used in the ecology community. The arrows indicate the alignment of slots that share the same constraint specification, i.e., the same semantic role. The corresponding semantic roles include the *OBJECT*, the *QUALITY*, and the *VALUE* that has been measured together with its *UNIT*. The slots carry the information that actually conveys the meaning of the weight measurement statement to a human reader. Blue arrows indicate slots with resources as values, and green arrows those with values. Slots with red borders indicate problems with terminological interoperability: OBO uses an instance of the class 'pato:weight', while OBOE, in this example, uses an instance of the class 'ncit:weight'. However, since 'pato:weight' and 'ncit:weight' are synonymous terms and can therefore be mapped, this is not a problem.

# What makes a term a good term and a schema a good schema?

First and foremost, a good schema for a (meta)data statement must cover all the information that needs to be documented, stored, and represented for the corresponding type of statement. However, beyond that, there are many other criteria for evaluating schemata. Most of these relate to the different operations one wants to perform on the (meta)data, and the formats required by the corresponding tools, which determine the **degree of machine-actionability** of the (meta)data. These include search operations (i.e., the findability in FAIR), but also reasoning and all kinds of data transformations, such as unit conversion for measurement data. Communicating with humans is another set of operations that needs to be considered when evaluating (meta)data schemata, as it relates to cognitive interoperability and thus the human-actionability of (meta)data.

Unfortunately, different operations are likely to have different requirements on a schema, and the tools that execute these operations may have their own requirements. For example, optimizing the findability of measurement data requires a different data schema than optimizing reasoning over them or their reusability. A given schema therefore needs to be evaluated in terms of the operations to be performed and the tools to be used on the (meta)data, often involving trade-offs between different operations that are prioritized differently in order to achieve an overall optimum. An example is the trade-off between reasoning and human-readability, as discussed above in the context of the dilemma between machine-actionable and human-actionable (meta)data schemata (Fig. 1).

As a consequence, for a given type of statement, there is likely to be a need for more than one corresponding schema. This is mainly because, besides historical reasons, different research communities often have different **frames of reference** and thus emphasize different aspects of a given type of entity, resulting in the need for different terms for the same type of entity (resulting in issues with ontological and thus terminological interoperability, but not necessarily with referential interoperability), but also because different research communities want to perform **different operations** on the (meta)data, different types of schemata. Since operations on (meta)data can be performed with different sets of tools, not only the structure of the schema is important, but also the format in which it can be communicated with such tools. For example, some tools require (meta)data to be in RDF/OWL, others in JSON, as CSV, or as a Python or Java data class.

Obviously, FAIRness is not sufficient as an indicator of high quality (meta)data—the use of (meta)data often depends on their **fitness-for-use**, i.e., their availability in appropriate formats that conform to established standards and protocols that allow their direct use, e.g., when a specific analysis software requires data in a specific format.

Therefore, although agreement on a common vocabulary and a common set of schemata would be a solution for semantic interoperability and machine-actionability of (meta)data across different research domains, this is unlikely to happen, and **we have to think pragmatically** and emphasize the need for ontological and referential term mappings and schema crosswalks for terminological schematic interoperability.

# The need for a machine-actionable Rosetta Stone for (meta)data that acts as an *interlingua* for specifying reference terms and reference schemata to support cognitive and semantic interoperability

Above, we discussed the role of terms and statement structures (i.e., syntax trees and (meta)data schemata) in reliably communicating the meaning and thus the semantic content of (meta)data statements. Statement structures specify syntactic positions or slots with semantic roles or constraint specifications for a given statement type. To achieve semantic interoperability, we therefore need controlled vocabularies (i.e., **ontologies**) and **ontological and referential term mappings** across ontologies for **FAIR terms** and their **terminological interoperability**. And we need **(meta)data schemata** and **ontological and referential schema crosswalks** for **FAIR (meta)data statements** and their **schematic interoperability**.

We also discussed why we think that it is impossible to agree on a best term for every possible type of entity and a best schema for every possible type of statement, due to varying frames of reference and operational priorities. Therefore, we think that we need something like a **machine-actionable Rosetta Stone** to support the establishment of semantic interoperability across different terms and different schemata for a given type of (meta)data statement. This Rosetta Stone needs to function like an ***interlingua***, with which term mappings and schema crosswalks can be easily specified and operationalized. The building blocks of the *interlingua* are **reference terms**, **reference datatype specifications**, and **reference schemata**. Each entity type must have specified a corresponding reference term, and each statement type must have a corresponding reference schema. Terms from controlled vocabularies can be mapped to their corresponding reference term, and schemata to their corresponding reference schema. Constraint specifications for slots of reference schemata must refer to reference terms in the case of resources, and to reference datatype specifications in the case of values. These three types of building blocks take over the role of **mediating connectors**, so that it would no longer be necessary to specify schema crosswalks for every possible pair of schemata of a given type of (meta)data statement and to specify term mappings for every possible pair of terms. This would minimize the number of schema crosswalks and term mappings that need to be specified in order to achieve schematic and terminological interoperability for a given type of statement (Fig. 4).

Ideally, a reference schema is based on a generic Rosetta modeling paradigm that allows the reconstruction of the natural language statement underlying the datum. At the same time, it should document this statement using a formalized structure to ensure its human- and machine-actionability. With respect to human-actionability, the Rosetta modeling paradigm should reflect as closely as possible the structure of natural language statements, favoring lean over complex models, with the aim of reducing overall modeling complexity and modeling burden. Many schemata are very complex and include positions with resources that do not directly align with any input slot (e.g., '*scalar measurement datum*' and '*scalar value specification*' in Fig. 2E). Such schemata are not suitable for use as reference schemata.

Schemata that conform to the Rosetta modeling paradigm should be easy to understand and to apply, allowing any producer of (meta)data to specify new reference schemata for types of statements that do not yet have a reference schema assigned to them, and allowing any application developer to readily use their (meta)data. It should not require experience in semantics and knowledge engineering on the part of the data producer and the application developer.

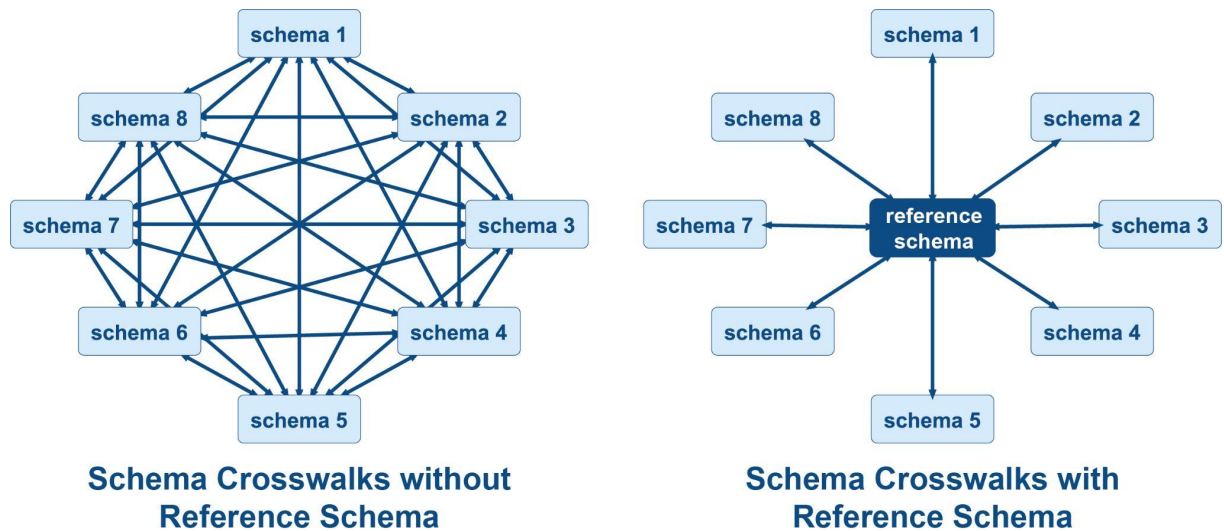

**Figure 4: Number of schema crosswalks required. Left)** The number of schema crosswalks required to achieve schematic interoperability between 8 different schemata is with 28 very high, because each possible pair of schemata has its own crosswalk. **Right)** With a reference schema playing the role of an intermediary schema, the number of required schema crosswalks can be reduced to a minimum of only 8, which significantly reduces the effort required to establish schematic interoperability for the corresponding type of (meta)data statement. The same is true for terms and term mappings.

With regard to the structure of reference schemata, we need to consider the machine-actionability of the resulting (meta)data statements. In other words, we need to consider which operations are important for such a reference schema. While reasoning is important for domain knowledge, especially when developing ontologies, other types of operations such as searching and exploring are more important in the context of empirical (meta)data and (meta)data management in general. However, regardless of the choice of operations and associated tools, the application of reference schemata must result in FAIR (meta)data that are machine-readable and machine-interpretable in order to be machine-actionable.

As for reference terms, they should ideally be collected in a large, controlled cross-domain vocabulary and should be machine-readable and machine-interpretable to be machine-actionable.

# Rosetta Stone and machine-readability: UPRIs, XML Schema datatypes, and RDF for communicating terms, datatypes, and statements

We have discussed above that a dataset documented in an ASCII file already qualifies as machine-readable. This readability includes all the terms and statements contained in the file. However, the ASCII file alone does not enable the machine to identify where exactly a term, a value, and a statement begin and end in a dataset. We would need to specify additional information in an application to enable a machine to make such distinctions.

Because we want to communicate (meta)data efficiently and reliably across machines and between machines and humans, and because we want to disambiguate terms, values, and statements, we suggest using **UPRIs** to communicate terms, **XML Schema datatypes** to communicate datatypes, and **RDF** to communicate statements. We use UPRIs for:

1. **instance-terms**: instance resource, including named individuals, that use proper names as their labels, refer to individual entities, and instantiate classes;
2. **class-terms**: class resources that use general/kind terms as their labels and refer to the extension of their class, i.e. the set of instance resources that instantiate the class. Class-terms can also refer to types of statements and their defining verbs or predicates.
3. **property-terms**: property resources that use verbs or predicates as their labels and refer to a specific type of action or attribute.

Machines are able to recognize unambiguously where a UPRI or a value (i.e., a literal) associated with an XML Schema datatype begins and where it ends. Both can be used in RDF to make statements using the RDF triple syntax of *Subject-Predicate-Object*, and a machine has no problem to recognize where a triple starts and where it ends.

Unfortunately, if you look at the predicate-argument-structure and compare the structure of triples with that of syntax trees, you will see that they are quite different and therefore do not properly align: the *Predicate* of a triple is always and necessarily **binary**, i.e., triples always have exactly one subject and one object argument. The predicates of natural language statements, however, are not necessarily always binary, as in the statement "*This tree* has part *an apple*", but **n-ary**, as in "*This apple* has *a weight* of 212.45 grams"; "*Anna* travels by *train* from *Berlin* to *Paris* on the *21st of April 2023*". Therefore, ontology properties (i.e., the *Predicate* resources used in triples) do not map in a one-to-one relation to natural language predicates, and we often need to model natural language statements using multiple triples (cf. Fig. 1 and Fig. 2E). This is possible in principle: If a triple uses a UPRI in its *Subject* position that is used by another triple in its *Object* position, these two triples combine to form a semantic graph, so we can use multiple triples to model an n-ary natural language statement. However, semantic modeling of n-ary statements in RDF is often challenging, significantly increasing the semantification burden and usually resulting in graphs that are overly complex for humans to comprehend.

In the following, we will stay with the RDF framework, as we believe it is the most flexible, allowing the specification of both ontologies and (meta)data schemata, as well as term mappings across ontologies and schema crosswalks across schemata. However, we will introduce a modeling paradigm that reflects the structure of English and provides a generic pattern for modeling n-ary statements in RDF. In all of this, we are trying to take a **pragmatic approach** that may not satisfy all the requirements for knowledge management that one would wish for in an ideal world, but from which we hope to achieve practical improvements.

# Rosetta Stone and machine-interpretability: Wikidata and a modeling paradigm for (meta)data statements based on English

In the search for reference terms, we turn to **ontologies and controlled vocabularies** because they provide terms that carry meaning through ontological definitions (and ideally also recognition criteria for designation and recognition tasks), assign a UPRI to each term, and organize the terms in a taxonomy. They, thus, provide **machine-interpretable terms**. With respect to cognitive interoperability, **Wikidata** lends itself as a prime candidate for a repository of reference terms: it covers many terms, especially in English, which is the de facto *lingua universalis* in science and academia. For many terms, it provides labels in multiple languages. It also covers many terms from different ontologies, and it provides an interface and workflows so that anyone can add more terms to it. Since each term in Wikidata has its own UPRI, the problem of homonymy of natural languages is solved, because even if two terms may have identical labels but differ in meaning and/or referent, they can still be distinguished by a machine because they have different UPRIs.

While terms carry meaning through their ontological definitions, statements carry meaning through their terms and the syntactic positions in which they are placed. As discussed above, not all natural language statements are based on predicates with a binary valence, so their predicates do not map directly to ontology properties. Because RDF is highly expressive, a given type of natural language statement can be modeled in RDF in many different ways. The core of our Rosetta approach is therefore a particular **modeling paradigm for types of statements**. To meet the requirements of cognitive interoperability, we follow the basic idea that this modeling paradigm should be as generic and simple as possible, reflecting as much as possible structures that we are already familiar with from a natural language like English. In addition, the paradigm must support the specification of new reference schemata so that it becomes a straightforward task that does not require any background in semantics—it should allow **the semantic modeling step to be automated**. This can only be achieved, if we come up with a **very generic structure for the model**. This structure must be applicable to any type of statement, regardless of its **n-aryness**. To be lean, it should store only what is necessary to recover the meaning of the statement, which should be the same as storing only the information a user needs to provide to create a new statement. For example, instead of creating the entire subgraph shown in Figure 2E, for a weight measurement statement, it should be sufficient to store only the resources for the measured object and the quality, together with the value and the unit, with the main focus on **always being able to reconstruct the original user input or data import for a given statement**. Another criterion that the generic model must meet is that it must facilitate the seamless derivation of queries from it. Each one of these criteria is important, because, in the end, the reference schemata must support semantically interoperable (meta)data statements with which not only machines but also humans can interact. But how do we get there?

For the time being, we limit ourselves to statements as the minimum information units, apart from terms. We can always extend the generic model later to create reference schemata that cover groups of (meta)data statements, and with semantic units (50) we have already provided a framework within which this should be straightforward.

Next, we need to abstract the structure of syntax trees from natural language statements to their syntactic positions and associated semantic roles. Since we do not need to be able to represent the full expressiveness of natural language statements when documenting (meta)data statements, we can restrict ourselves to statements with a rather simple structure of subject, transitive verb or predicate, and a number of objects. In this first attempt to develop a machine-actionable Rosetta Stone, we do not consider passive forms, no tenses, and we do not need to distinguish between different possible syntactic alternations in which a verb or predicate can express its arguments. **The generic model underlying our modeling paradigm is thus similar to a highly simplified frameset** (see discussion above), **specifying a subject-position and a number of required and optional object-positions, each with its associated semantic role in the form of a thematic label and a corresponding constraint specification**. Its structure is thus an abstraction of the structure of a syntax tree.

Different types of statements can be distinguished on the basis of their underlying predicates (i.e., relations), resulting in a **predicate-based classification of types of statements**. For example, the apple weight measurement statement from Fig. 2A is of the type *weight measurement statement*.

As described above, following the **predicate-argument-structure**, each predicate has a **valence** that determines the number and types of **arguments** that it requires to complete its meaning. For example, a has-part statement requires two arguments—a subject representing the entity that has some part and an object representing the part—resulting in a statement that can be modeled as a single triple following conventional triple modeling schemes, e.g., 'SUBJECT *has part* OBJECT'. Without the specification of a subject and an object, the has-part predicate cannot complete its meaning. **Adjuncts** can additionally relate to the predicate, but they are not necessary to complete the meaning of the predicate but provide optional information, such as a timestamp specification in the has-part statement, e.g., 'SUBJECT *has part* OBJECT *at* TIMESTAMP', which would require multiple triples for modeling in RDF. Subject and object phrases are the most common arguments and adjuncts. So we can say that each statement relates a resource, which is the subject argument of its predicate, to one or more literals or resources, which are the object arguments and adjuncts of its predicate. The subject argument is what we call the **subject** of the statement, and the object argument(s) and adjunct(s) are its **required and optional object(s)**. Every statement has such a subject and one or more objects.

According to this model, we can distinguish different types of predicate by the total number of subjects and objects they relate within a given statement. A statement like '*Sarah* met *Bob*' is a statement with a **binary relation**, where we refer to '*Sarah*' as the subject of the statement and '*Bob*' as its object. If we add a date to the statement, such as in '*Sarah* met *Bob* on *4th of July 2021*', it becomes a **ternary relation** with two objects[3]. If we add a place, it even becomes a **quaternary relation**, as in '*Sarah* met *Bob* on *4th of July 2021* in *New York City*'. This is open-ended in principle, although it is limited by the dimensionality of the human reader's ability to comprehend n-ary relations[4]. Regardless of this limitation, statements can be distinguished into binary, ternary, quaternary, and so on, based on the number of subjects and objects that their underlying predicates relate to. Furthermore, based on the distinction between arguments and adjuncts, we can distinguish **objects** that are necessary and thus **required** to complete the meaning of the statement's predicate from **objects** that are **optional**.

Looking again at the example above, if we were to model the statement '*Sarah* met *Bob* on *4th of July 2021*' in a knowledge graph, the objects '*Bob*' and '*4th of July 2021*' would be modeled differently. Whereas '*Bob*' is likely to be modeled as a resource that instantiates a class 'person' (wikidata:Q215627), '*4th of July 2021*' is likely to be modeled as a literal associated with the datatype xsd:date. Therefore, in addition to distinguishing arguments and adjuncts, each with their associated semantic roles and thematic labels, one can distinguish objects by their type into resources via their respective UPRIs and literals and specify **class/datatype constraints based on their associated semantic roles**. Resources, in turn, can be either named-individuals, classes, or properties (and, when following the semantic unit framework, some-instance and every-instance resources as well (50)). We shall refer to them as **resource-objects** and to literal-based objects as **literal-objects**. Statements can be distinguished by the number of resource-objects and the number of literal-objects they contain. With resource-subjects, resource-objects, and literal-objects, we now have the different elements that each reference schema must cover. The next step is to work out how best to relate them to each other and to the statement.

---

[3] Many properties of the Basic Formal Ontology 2.0 are actually ternary relations because they are time-dependent (51). For example, "*subject* located in *object_A* at *t*".

[4] Humans can hold only 5-9 items in memory (52).

**The light version of the Rosetta approach**

The Rosetta modeling approach needs to relate the subject-resource to the different types of object-resources and object-literals of a given type of (meta)data statement. To avoid the problems of modeling statements that have n-ary predicates, and to reflect as closely as possible the structure of simple natural language statements in English that consist of only one verb or predicate, we classify all types of statements based on their predicate, and use instances of the respective classes to link the subject and object-resources, as well as the object-literals. So the statement '*This apple has a weight of 212.45 grams*' would instantiate a '*weight measurement statement*' class, and the corresponding reference schema would link an instance of '*apple*' (wikidata: Q89) as the statement's subject-resource via a '*has subject*' property to an instance of this statement class. The schema would also require two additional arguments to be added: (i) a value of 212.45 with datatype xsd:float as the object-literal and (ii) a named-individual resource '*gram*' (wikidata:Q41803) as the object-resource. The schema links the statement instance resource to these object-arguments via a '*required object position*' property (Fig. 5). Comparing this schema with the weight measurement schemata from OBO and OBOE (cf. Fig. 3), it is immediately apparent that, on the one hand, fewer triples are required to model the statement—i.e., three instead of five or six— and on the other hand, much fewer classes are required. The reference schema is simpler and contains only input slots and no additional positions such as '*scalar measurement datum*' and '*scalar value specification*' in the OBO schema or '*observation*' and '*measurement*' in the OBOE schema. A human reader is not interested in these additional positions and their resources—they only want to see the information from the input slots.

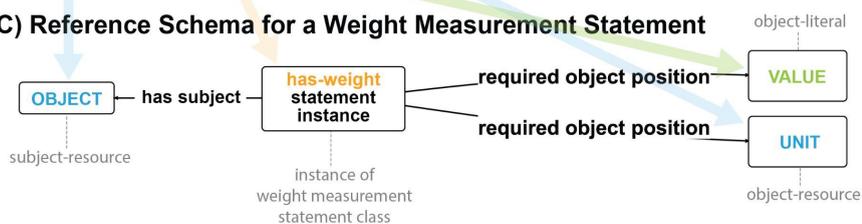

Figure 5: From the structure of a natural language statement to the structure of a reference schema. A) A natural language statement with the predicate *has_a_weight*. B) The corresponding formalized statement, with the syntactic positions and their associated semantic roles highlighted in color. C) The reference schema for the weight measurement statement, following the Rosetta modeling paradigm.

The same modeling approach can be applied to any simple English statement consisting of a single verb or predicate. We therefore chose this modelling approach as the Rosetta modeling paradigm for reference schemata (see Figure 6).

As each instance of such a statement class represents the statement as a whole, including its verb or predicate, one can use this resource to make statements about that statement, including about (i) the provenance of the statement, such as creator, creation date, curator, imported from, etc., (ii) the UPRI of the reference schema that the statement instantiates, (iii) the copyright license for the statement, (iv) access/reading restrictions for specific user roles and rights for the statement, (v) whether the statement can be edited and by whom, (vi) a specification of the confidence level of the statement, which is very important, especially in the scientific context (53,54), where lack of it can cause problems such as citation distortion (55), (vii) a specification of the time interval for which the statement is valid, and (viii) references as source evidence for the statement, to name some possibilities. In other words, following the Rosetta modeling paradigm, one always gets statements, each represented by its own dedicated resource, so that one can make statements about each of these statements without having to apply RDF reification (56) or RDF-star (57,58), which are feasible for referring to individual triples but not for larger subgraphs such as a measurement datum with a 95% confidence interval (see Fig. 1, middle), for which the latter two approaches are inefficient and complicated to query. Named Graphs seem to be another solution for such larger subgraphs (56), and one could always organize all triples belonging to a statement into their own Named Graph using the UPRI of the statement instance resource as the UPRI of the Named Graph.

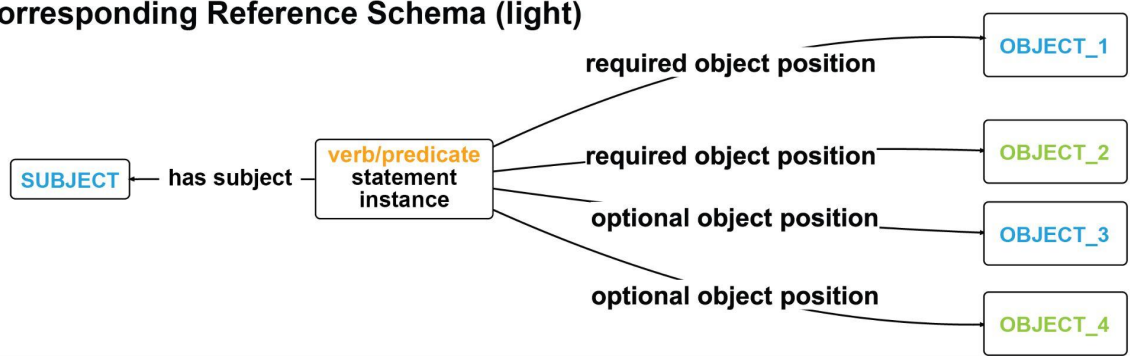

**Figure 6: From a formalized natural language statement to the corresponding reference schema according to the light version of the Rosetta approach. A)** A formalized statement with its syntactic positions and associated semantic roles highlighted in color. **B)** The reference schema for the statement from A), following the modeling paradigm of the light version of the Rosetta approach. The statement instance resource indirectly indicates the verb or predicate of the statement, shown in orange. Object arguments ('*required object position*') and adjuncts ('*optional object position*') can be either object-resources (in blue) or object-literals (in green).

Each argument in a given reference schema can be understood as a particular syntactic position, which we model in the schema as a slot, for which we can specify the corresponding semantic role in the form of a constraint specification—either as XML Schema datatype specification for an object-literal, which can be supplemented with a specific pattern or range constraint, or as a Wikidata class specification for a subject or an object-resource, which restricts the type of resources that can be located in a particular slot to that class or any of its subclasses. Corresponding reference schemata can be specified as **SHACL shapes**, for example. Statements modeled according to the same shape are **machine-interpretable statements**.

A given reference schema can be extended to include more object adjuncts. Adding new object adjuncts to a reference schema is not problematic because any statement instance that was created using an older version of the schema will still be compatible with the new schema, since object adjuncts are only optional objects and thus are not required to comply with a reference schema.

**The full version of the Rosetta approach, supporting versioning and the tracking of an editing history**

Some knowledge graphs are rapidly evolving, and their content is the result of collaborative or even crowdsourced editing, where any user can edit any statement in the graph, including statements created by other users. For such knowledge graphs in particular, it may be important to be able to track the editing history at the level of individual statements. And if the knowledge graph were to allow its users to cite one or more of its statements, thus making the knowledge graph a valuable resource for scholarly communication, it would also require a versioning mechanism that provides a citable resource for sustaining the cited content. With such a mechanism, the knowledge graph can evolve continuously through user input, while still being citable. The statement versioning mechanism of the full version of the Rosetta approach supports this, and it supports tracking the editing history for each individual statement and each particular object-position. To support this, the Rosetta modeling paradigm must be adapted.

In the modeling paradigm of the full version of the Rosetta approach (Fig. 7), only the subject-resource is directly linked to the statement instance. The object-resources and object-literals are only indirectly linked to the statement instance through instances of corresponding object-position classes. Each reference schema has an object-position class defined for each of its object arguments and adjuncts, so that each particular statement of a given statement type has, in addition to an instance of the statement class, an instance of each object-position class, to which the actual object-resources and object-literals are linked. The number of object-position classes that a given reference schema distinguishes depends on the n-aryness of the underlying statement type. The dependency of object-position classes on their corresponding statement is also documented at the class level: Each statement class links in its class axioms to its corresponding required and optional object-position classes.

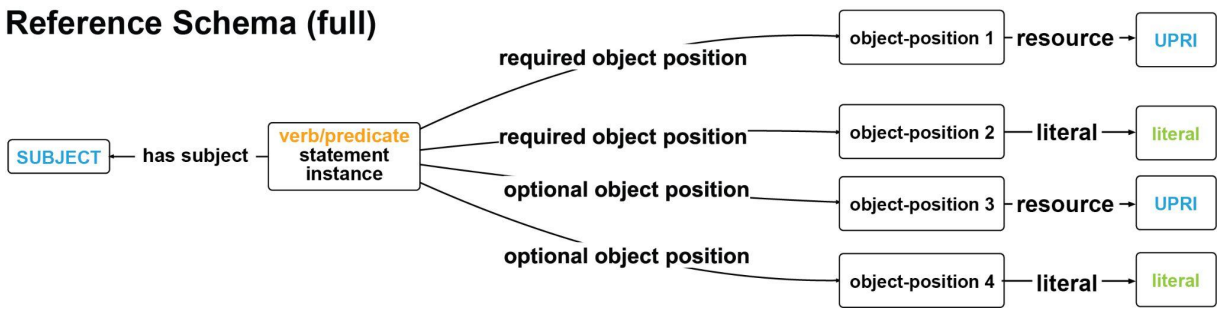

**Figure 7: Structure of a reference schema following the full version of the Rosetta approach.** The reference schema for the statement from Figure 6 A), following the modeling paradigm of the full version of the Rosetta approach. Compared to the light version of the approach (Fig. 6B), the full version defines for each statement class, depending on the aryness of the statement, one or more accompanying object-position classes—one for each object argument and adjunct. For a given statement, the corresponding object-position classes are instantiated and linked to the corresponding statement instance, depending on whether they are arguments ('*required object position*') or adjuncts ('*optional object position*'). The actual object-resources (blue) and object-literals (green) are linked to their respective object-position instance. This structure supports the versioning of statements and the tracking of the editing history for each object position of each statement in a knowledge graph.

Statements modeled according to the full version of the Rosetta modeling paradigm are downward compatible with those modeled using the light version, and can therefore be transferred to them. The reverse is not possible, as it requires the specification of corresponding object-position classes that are not required in the light version.

In addition to the object-resource or object-literal, further information can be associated with and thus documented for each object-position instance, such as provenance metadata about the particular input event (e.g., creator, creation date, imported from). Thus, one can think of an object-position instance of a particular statement to document a particular input event associated with that position. One can also specify that a particular **logical property** of the verb or predicate of the statement applies to a particular object-position by using a corresponding Boolean annotation property (e.g., '*transitive*') with the object-position instance. This way, you can, for example, document that the transitivity of a has-part statement applies to the resource specified for the PART object-position.

The mechanism for statement versioning and for tracking the editing history uses these structural changes in the reference schema according to two basic ideas:

**1) Soft-delete: no input can be deleted or modified after it has been added to the graph.** Each editing step results in the addition of new triples to the graph of a corresponding statement, with the information provided being linked to a corresponding newly added object-position instance. In other words, the object-position instances belonging to a particular statement are not updated, but each editing step adds new instances to the graph. To identify newly added object-position instances, a Boolean property is used on the object-position instance and set to '*true*'. For example, if a user updated the VALUE position of a weight measurement statement, the VALUE position instance that was previously associated with the last added value, and therefore the one with the value '*true*' for the '*current*' property, would be set to '*false*', and a new VALUE position instance with the '*current*' value '*true*' would be added, and its information displayed when a user navigates to the statement.

Each object-position instance added in this way instantiates the corresponding object-position class, so that a query can identify all instances of the same object-position of a given statement. All previously added object-position instances of the same object-position class are still in the knowledge graph and linked to the corresponding statement instance, and their information could be accessed at any time and sorted by their creation date if required, thus providing a **detailed editing history** for each object-position of a statement, but also for the statement as a whole. In the case of the weight measurement statement example from above, information about who entered when which weight measurement value could be presented in the user interface (UI) of a knowledge graph. Each object-position holds this information. This information is valuable in the context of managing collaborative efforts to edit content in a knowledge graph, and also for crowdsourced knowledge graphs to identify who added/edited what and when.

The same approach is applied not only to individual object-positions, but also to individual statements. The respective statement instance also indicates its status through the '*current*' property. If a user wants to delete a statement, instead of actually **deleting** the triples associated with the statement, the statement instance is set to '*false*' via the '*current*' property. By default, the application does not display information about resources that are set to '*false*', and in the UI it looks as if the information has been deleted, although it is still available in the knowledge graph and all associated provenance metadata are still accessible (fulfilling principle A2. of FAIR).

**2) A user can create a version of any statement in the knowledge graph, which then cannot be edited and is therefore persistent over time.** Users can create versions of specific statements. Each such version is represented by its own node in the graph. It has its own UPRI, which could take the form of a DOI. Consequently, such versions can be referenced and cited in publications. The version node also tracks metadata through its properties, like any other statement. Each version could,

additionally, also have a **content identifier (CID)** that uses cryptographic hashing for **decentralized content addressing**. There may be several versions of a given statement. The node representing the latest version is linked to its statement instance by the '*hasCurrentVersion*' (PAV:hasCurrentVersion) property, while all other version nodes are linked to it by the '*hasVersion*' (PAV:hasVersion) property. The version nodes of a statement are linked to each other by a chain of '*previousVersion*' (PAV:previousVersion) properties, starting with the latest version and going back to its respective previous versions.

All object-position instances belonging to the versioned statement are also updated to include the UPRI of the versioned statement as a value for their respective '*versionID*' property. With this information, it is always possible to access the data belonging to that version by querying all object-position instances with the UPRI as a value for this property. The '*versionID*' property can be applied multiple times to the same object-position instance to track more than one version UPRI. This is necessary, because not all object-positions of a given statement may have been edited between two versions, and thus a particular object-position instance may belong to more than one version.

Both aspects, the soft-delete and the versioning, have been demonstrated in a small Python-based prototype of a FAIR scholarly knowledge graph application that employs semantic units and knowledge graph building blocks and that uses Neo4j as the persistence-layer technology (62). It includes versioning of semantic units and automatic tracking of their editing history and provenance. The prototype is available at https://github.com/LarsVogt/Knowledge-Graph-Building-Blocks.

**The Rosetta Framework primarily models statements rather than some human-independent reality**

It is important to note that **the Rosetta Framework does not claim to model and represent a human-independent reality**, as other approaches to semantic modeling attempt to do, such as the Open Biological and Biomedical Ontology (OBO) Foundry, with the Basic Formal Ontology (BFO) (51) as its top-level ontology. Instead, it follows a **pragmatic approach with a focus on the efficient and reliable communication of information of all kinds** between humans and machines and across machines, including but not restricted to (meta)data statements. For the time being, the Framework is limited to terms and statements as meaning-carrying units of information, but can be extended to larger units in the future (see also the concept of semantic units to which it can be easily adapted (50)).

As each statement is represented in the graph with its own resource and as an instance of a particular statement class, it is straightforward to make statements about these statements. This includes, in addition to the classification of statements based on their verb or predicate already discussed, the ability to further classify any given statement as an instance of other types of classes, allowing sophisticated classification of statements across multiple independent classification criteria.

For example, one can distinguish between different types of statements regarding their truth function or reference. When we talk about the referent of a term that is a proper name, we are referring to a particular entity that is known by that name, and when the term is a kind term, we are referring to the set of entities that are instances and thus referents of the respective class. With statements, we can do something similar with respect to the subject of the statement, and we can distinguish the following types of statements (for a discussion, see also (50)): (1) The referent of an **assertional statement** is the particular known individual entity (i.e., named individual) that takes the subject-position of the statement. Assertional statements are taken to be true for that particular individual entity. Instance graphs and ABox expressions are examples of assertional statements (60).

(2) The referent of a **contingent statement** is some instance of a class which takes the subject-position of the statement. Contingent statements are taken to be true for at least one instance of that class. (3) The referents of a **prototypical statement** are most instances of a particular class, indicating that the statement is true for most, but not necessarily all, instances of the class. As such, prototypical statements are a special kind of contingent statement. (4) The referent of a **universal statement** are all instances of a class. A universal statement is necessarily true for every instance of that class. Class axioms of ontology classes, and thus TBox expressions, are examples of universal statements (60).

Another possibility is to classify a given statement instance as an instance of the '*negation*' class, indicating that its semantic content is negated. This can be used to express typical negations, but also absence statements, which are often needed when describing specific objects, situations, or events. In OWL, following the Open World assumption, modeling negations, including absences such as '*This head has no antenna*', requires the specification of appropriate class axioms and blank nodes. By classifying statement resources as instances of a class '*negation*', one can even model assertional, contingent, and prototypical statements with negations without having to model them as TBox expressions. The same applies to statements with cardinality constraints (e.g., '*This head has exactly 3 eyes*'). Modeling negations and cardinality constraints in this way is much simpler than modeling them as TBox expressions, and would thus increase their cognitive interoperability (see (50) for more details).

Many more classification criteria can be applied, leading to a sophisticated classification of different types of statements based on their meaning, their epistemic value and function, their referents, and their context (e.g., assumptions, ontological definitions, epistemological recognition criteria, questions, answers, logical arguments, commands/instructions, descriptions, hypotheses, knowledge, known facts).

**A Rosetta Editor for creating reference schemata**

The use of the light and full versions of the Rosetta modeling paradigm results in reference schemata that are easily understood by humans because the modeling paradigm reflects the structure of natural languages such as English. In addition, specifying reference schemata for new types of statements is not as demanding when following this modeling paradigm. Not only developers, but also researchers (and anyone else) who want to use knowledge graph applications and who are experts in domains other than semantics and knowledge modeling in particular, or computer science in general, will be able to specify new reference schemata if supported by an intuitively usable low-code Rosetta Editor. With the Editor, the need to develop semantic models to establish FAIR (meta)data within the RDF framework will no longer be a barrier (i.e., the semantification burden), and we expect that related tools will be easier to adapt, increasing the overall cognitive interoperability of the RDF framework for developers and users alike.

When specifying a new reference schema, users of the Editor will be guided through a list of questions, the answers to which allow the Editor to create the reference schema and its associated statement class (and object-position classes), **without having to make any semantic modeling decisions**:

1. **Provide some example statements.**
   Example statements can be informative for users to understand what type of statement the reference schema is modeling, and what types of statements can be covered by it. In addition, for Large Language Models, example statements can be used to assist with the following questions by providing possible answers which only need to be checked for correctness by the user.

2. **What is the predicate or verb of the statement?**
   The answer will be used to create the statement class label.

3. **Give a description or characterization of the type of statement you want to add. What kind of statement will it cover?**
   Provides a human-readable definition of the newly created statement class. It could also take the form of a definition of the predicate of the statement.

4. **Indicate the number of object-positions the statement should cover.**

5. **Give the subject-position and each object-position a short and meaningful label.**
   Each label acts as a **thematic label**, characterizing the **semantic role** associated with each position (cf. (45)). It is used for various purposes, including as a placeholder label for input fields and as a placeholder label for variables for the textual and graphical display templates (see below). It also provides the label for the corresponding object-position class for the full version.

6. **Specify which of these object-positions are required (i.e., arguments) to form a semantically meaningful statement with the subject and the predicate, with all other object-positions being optional (i.e., adjuncts).**
   The '*requiredObjectPosition*' property is used to specify which object-positions the statement instance will bind to, while the '*optionalObjectPosition*' property is used to bind to all other object-positions. In the Editor UI, this could be done using checkboxes.

7. **For each object-position, provide a brief description of the type of objects covered by the object-position and give some typical examples.**
   This is the human-readable definition for the corresponding object-position class. This text can also be used as a tooltip for the corresponding input field for adding or editing corresponding statements.

8. **For each object-position, decide whether the object must be represented in the form of a resource or a literal. In the case of a resource, choose a Wikidata class that best specifies the type of entity that is allowed for the object-position, and in the case of a literal, choose the datatype and define datatype-specific constraints if necessary.**
   This provides input constraints for each object-position, which are documented in the shape specification of the reference schema.

9. **Decide whether any logical properties apply to the predicate of the statement (e.g., transitivity) and, if so, which object-position is affected by them.**
   This is only optional, but required if you want to be able to automatically translate the statement graph resulting from instantiating a reference schema into an OWL graph over which reasoning can be applied. The Editor allows this information to be added to the relevant object-position class via an appropriate Boolean annotation property (e.g., '*transitive*').

10. **Write a human-readable statement using the thematic labels for the subject, the different object-positions, and the predicate.**
    Provides the dynamic label pattern for this type of statement (see display templates below).

The Rosetta Editor will process the input and create a reference schema along with a statement class and, if the full version of the Rosetta approach is followed, any required object-position classes. The schema itself could be specified in a YAML file, following the notation of **[LinkML](#)**, which can be translated into, e.g., [SHACL](#) shapes. The schema specifies slots/attributes with constraints that can be used by the frontend and a knowledge graph application for validation and input control purposes. Constraints include specifying datatypes as ranges such as '*xsd:float*', max-min, or pattern constraints, and class constraints for resources. Slots belonging to required object-positions would be declared as '*required: true*', indicating that this slot must have a value. [LinkML](#) is a general-purpose, platform-agnostic, object-oriented data modeling language and framework that aims to bring Semantic Web standards to the masses and that makes it easy to store, validate, and distribute FAIR data (61). It can be used with knowledge graph applications to schematize a wide variety of data types, from simple flat checklist standards to complex interrelated normalized data using inheritance. LinkML fits nicely into frameworks common to most developers and database engineers, such as JSON files, relational databases, document stores, and Python object models, while providing a solid semantic foundation by mapping all elements to RDF U(P)RIs.

LinkML also comes with several tools. Generators provide automatic translation from the YAML schema into various formats, including JSON schema, JSON-LD/RDF, SPARQL, OWL, SQL DDL, ShEx, GraphQL, Python data classes, Markdown, and UML diagrams. Loaders and dumpers convert instances of a schema between these formats. Generators thus allow integration with tools provided by other technical stacks. The SPARQL generator allows generating a set of SPARQL queries to be generated from a schema, and the Excel Spreadsheet Generator allows a spreadsheet representation of a schema to be generated. LinkML currently supports four different data validation strategies: 1) validation via Python object instantiation; 2) validation via JSON schema; 3) validation of triples in a triple store or RDF file via generation of SPARQL constraints; or 4) validation of RDF via generation of ShEx or SHACL.

# Rosetta Stone and semantic interoperability: Specifying term mappings and schema crosswalks

Different controlled vocabularies and ontologies may contain terms that have the same meaning and referent, and are therefore strict synonyms. Unfortunately, if their UPRIs differ, a machine will not be able to recognize them as synonyms, and statements using such terms will not be interoperable because the terms they use are not **terminologically interoperable** (10). The problem of the lack of terminological interoperability can be solved by specifying ontological term mappings for all strictly synonymous terms across different vocabularies and ontologies, as discussed above. Unfortunately, however, terms that appear to be strictly synonymous at first glance sometimes turn out not to be strictly synonymous on closer inspection, or the evaluation of their synonymy involves evaluating chains of interdependencies between numerous terms across different ontologies and is practically not feasible. In such cases, but also when terms refer to the same entities but differ in their ontological definitions, referential term mappings can be specified.

In the first implementation of the Rosetta Framework, we suggest being pragmatic and focusing on referential interoperability of terms and thus referential term mappings—ontological term mappings can still be added later. We plan to use Wikidata terms as referent terms for referential term mappings. The mappings and should be stored and made openly accessible and usable in a terminology service such as the TIB Terminology Service.

Analogous to referential term mappings, referential schema crosswalks will be specified to establish **schematic interoperability** between all schemata that model the same type of statement (10). With schema crosswalks specified via the appropriate Rosetta Framework reference schema (cf. Fig. 4), researchers can use a schema that is optimized for the set of operations and tools relevant to their particular project and research topic, while making the (meta)data they create referentially and schematically interoperable with all other statements created with schemata for which schema crosswalks have been specified for the same reference schema (Fig. 8).

All provenance and metadata statements associated with a particular dataset or a given data statement can be mapped to any provenance and metadata schema or model, such as PROV-O, PAV ontology (62), Dublin Core Metadata Initiative, or DataCite Schema, if a corresponding reference schema is specified along with the required **metadata crosswalks**.

In addition to specifying schema crosswalks to different (meta)data graph schemata, they can also be specified to different formats such as RDF/OWL, GraphQL, Python or Java data classes, JSON, and CSV. Since all of these formats must provide data slots for a given statement type that map to its positions and their associated semantic roles, mapping to non-graph-based formats should be analogous to mapping to graph-based formats (e.g., Fig. 2D).

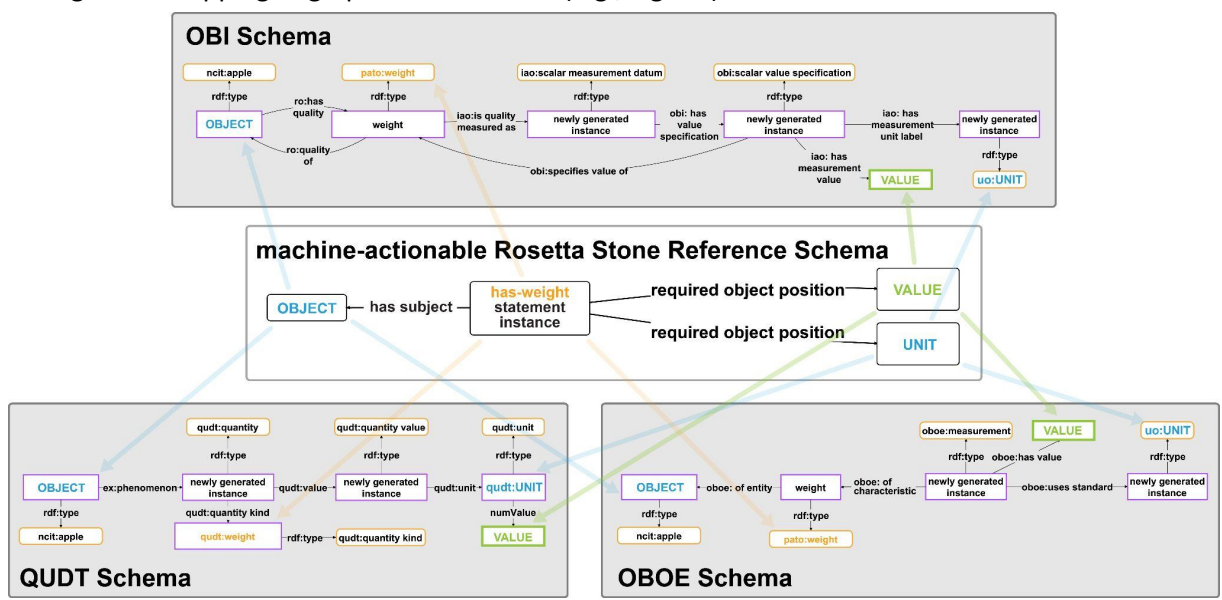

**Figure 8: Schema crosswalks from the Rosetta Stone reference schema for a weight measurement statement to three other schemata**. **Middle:** The Rosetta Stone reference schema for a weight measurement statement. **Top:** The OBI schema. **Bottom left:** The schema of the Quantity, Units, Dimensions, and Types Ontology (QUDT) (model transferred from (63)). **Bottom right:** The OBOE schema. Blue arrows indicate the alignment of object-resource slots, green arrows of object-literal slots, and orange arrows of the verb or predicate slots. When operationalizing crosswalks, term mappings are often required to satisfy the constraint specifications of the target schema. For example, if the OBI schema requires a class resource from the Units of Measurement Ontology (UO) for its UNIT slot, the Wikidata resource of the source must be translated into a corresponding UO resource, while for the QUDT schema it would have to be translated into a corresponding QUDT resource.

The Rosetta Editor must allow the specification of new graph-based and tabular (meta)data schemata and formats whenever needed, thus allowing adaption to newly emerging (meta)data formats and standards[5]. This takes into account the observation that FAIRness is not sufficient as an indicator of high (meta)data quality—the use of (meta)data often depends on its **fitness-for-use**, i.e., data must be available in appropriate formats that comply with established standards and protocols that allow their direct use, e.g., when a specific analysis software requires data in a specific format.

---

[5] In the case of new formats, this may require adjustments to the Editor.

With the Rosetta Framework and its Editor, any domain expert can begin to specify their own small **Rosetta module** for each type of statement that is relevant to them. The module comprises the statement class, a reference schema, and schema crosswalks. Rosetta modules facilitate the production of FAIR machine-actionable (meta)data statements. Using the Editor, any number of additional schemata can be specified at any time and linked to the reference schema through a schema crosswalk. These additional schemata can be used by applications to make their (meta)data, which are stored according to reference schemata, accessible in any other schema and format—either virtually through their UI or as an export option. As a result, (meta)data managed by an application that applies the Rosetta Framework are **decoupled from the application's storage model and can be readily used in other frameworks**.

And since schema crosswalks allow (meta)data conversion in both directions, they can also be used to import (meta)data from any schema into the reference schema. Combined with the concept of semantic units (50), the Rosetta modules resemble **Knowledge Graph Building Blocks**, i.e., small information modules for knowledge processing (for a discussion of Knowledge Graph Building Blocks, see (64)).

The ability to specify schema crosswalks that convert, for example, weight measurement statements that comply with a corresponding reference schema into data graphs that comply with the corresponding OBI schema also opens up the possibility for knowledge graph applications to establish workflows in which statements that meet certain criteria, such as having a certain confidence level or having a documented reference to a relevant source of evidence for the statement, are then converted into data graphs that comply with the OBI schema for weight measurements, thereby converting information from a schema that models statements into a schema that models a human-independent reality.

**From reference schemata to OWL-based schemata that allow reasoning**

(Meta)data statements and their corresponding statement classes can be understood as a formalized approach to modeling n-ary predicates. Each statement class refers to an n-ary predicate, and the statement class can be understood as an attempt to model this predicate as an ontology class instead of an ontology property. As a consequence, reasoning over property axioms such as transitivity or domain and range specifications is not straightforward with (meta)data statements resulting from instantiations of reference schemata, and tools established for OWL-based frameworks cannot be readily reused in the Rosetta Framework. Therefore, it is important for the Framework to provide **interoperability with OWL and description logics** by providing, by default, a corresponding **OWL-based schema** for each reference schema and its corresponding statement class. When a new reference schema is specified for a new type of statement, the Rosetta Editor uses the information provided for the newly specified statement class and for the reference schema to automatically specify an OWL-based schema and the associated schema crosswalk.

When we take a material has-part statement with a material object as the subject and another material object as the object argument, the corresponding reference schema takes on the structure shown in Figure 9, top. We can define a new object property '*has material part*' as a subproperty of '*required object position*', with the domain and range specified as 'MATERIAL OBJECT' (Fig. 9, bottom). An annotation property indicates that the '*has material part*' property belongs to the '*material has-part statement*' class. The domain and range specifications are taken from the constraints of the two slots that the property connects, and any logical property axioms, such as transitivity, are taken from the corresponding specifications on the range object-position class (see discussion above).

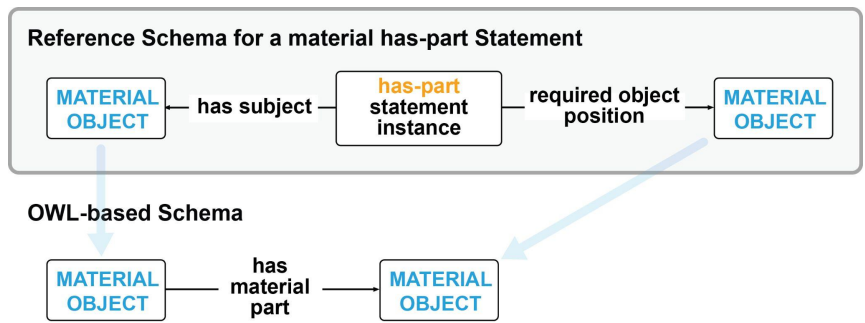

Figure 9: Schema crosswalk between the Rosetta Stone reference schema of a material has-part statement (top) and a corresponding OWL-based schema (bottom). Blue arrows indicate the alignment of the slots.

# Rosetta Stone and cognitive interoperability: Specifying display templates and using a query builder

**Display templates**

Humans usually do not want to see the (meta)data of a knowledge graph in the form of triples—they do not want to read them in any of the RDF serializations, nor do they want to visualize them as a graph. In order to display the content of a (meta)data statement modeled according to the Rosetta approach in a human-actionable way, a frontend application needs a **display template** that specifies how a rendering function can translate a given (meta)data structure into a human-readable statement. Display templates organize information from the (meta)data structure along with additional prepositions to be presented in the UI.

For example, if a weight measurement with a 95% confidence interval were stored according to an appropriate reference schema, the schema would specify four literal-object-positions (*MAIN_VALUE*, *UPPER_VALUE*, *LOWER_VALUE*, *INTERVAL_VALUE*) and one resource-object-position (*UNIT*). The form-based display template could specify that this information should be displayed in the frontend as '*SUBJECT* has weight (*INTERVAL_VALUE*% conf. interval): *MAIN_VALUE* (*LOWER_VALUE-UPPER_VALUE*) *UNIT*'. Another example is a travel statement where the corresponding reference schema specifies one required resource-object-position (*DESTINATION_LOCATION*), two optional resource-object-positions (*DEPARTURE_LOCATION*, *TRANSPORTATION*), and one optional literal-object-position (*DATETIME*). The corresponding display template would display a travel statement in the frontend as '*PERSON* travels by *TRANSPORTATION* from *DEPARTURE_LOCATION* to *DESTINATION_LOCATION* on the *DATETIME*'. In other words, the subject-position and the various object-positions (i.e., the syntactic positions with their associated semantic roles) are mapped to corresponding variables within a string to form a human-readable statement (see Fig. 10, top). We call such **textual** display templates **dynamic labels**.

In addition to textual display templates, it is also possible to specify graphical display templates for a **mind-map-like representation of a statement** using **dynamic mind-map patterns**. Dynamic mind-map patterns use a label for the predicate underlying the corresponding statement type and, if there is more than one object-position, labels for relating the various objects to the predicate. As a result, the statement can be visualized as a mind-map like graph, where each subject and object is represented as a node with the label of the corresponding resource from the underlying graph of the (meta)data statement (see Fig. 10, bottom). Such graphical representations of statements can also be combined to form a mind-map of larger contexts and interrelationships that connect the dynamic mind-map patterns of multiple statements. Mind-map-like representations of complex interrelationships between different entities are easier to understand than form-based textual representations, thus increasing the human-actionability of a knowledge graph.

**Dynamic Label (Textual Display)**

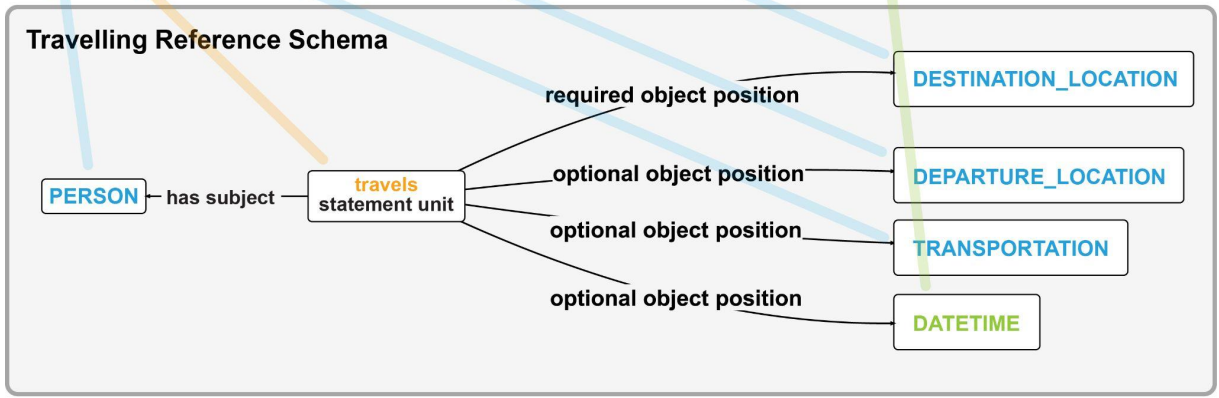

**Dynamic Mind-Map Pattern (Graphical Display)**

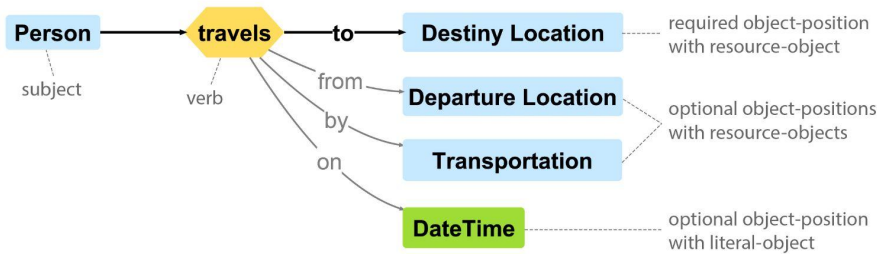

**Figure 10: Textual and graphical displays of a statement based on its reference schema. Middle:** The reference schema of a traveling statement. **Top:** A textual display of the traveling statement, i.e., a dynamic label that is associated with the reference schema. Note how the subject-position and the object-positions from the reference schema align (arrows) to variable-positions (i.e., positions with associated semantic roles) in the dynamic label template. **Bottom:** A graphical display of the traveling statement, i.e., a dynamic mind-map pattern that is associated with the reference schema. The alignment of the subject-position and the object-positions from the reference schema to nodes in the dynamic mind-map pattern is similar to the alignment for the dynamic label template (arrows not shown here for clarity).

(Meta)data can be communicated to the presentation layer in the UI, with the templates filtering the complex data structure for the information relevant to a human user, using **dynamic labels and dynamic mind-map patterns** to present statements, **decoupling human-readable data display from machine-actionable data storage**.

A given reference schema can have multiple dynamic labels and dynamic mind-map patterns associated with it. It can be beneficial to be able to choose between different display templates depending on the context in which the (meta)data are accessed (PC vs. smartphone, expert user vs. layperson, etc.). The specifications of each display template should be associated with its corresponding reference schema and statement class, thus contributing to the easily accessible and usable Rosetta module mentioned above, moving it even closer to the concept of **Knowledge Graph Building Blocks** (64).

**A Rosetta Query-Builder**

We believe that **findability** is the most important aspect of any tool that stores and documents (meta)data. If relevant (meta)data cannot be found, no other operations can be performed on it, and interoperability issues become secondary. In the context of knowledge graphs, specific query endpoints have been established along with corresponding graph query languages to interact with the graph. These are [SPARQL](SPARQL) for RDF- and OWL-based knowledge graphs and [Cypher](Cypher) for labeled property graphs such as [Neo4J](Neo4J). Our personal experience is that most users and software developers have no experience with graph-based databases, are not familiar with graph query languages and their benefits, and therefore do not see the need to learn them. And even those who are familiar with them report that writing more complex queries can be challenging. **Apparently, the need to write SPARQL or Cypher queries is a barrier to interacting with knowledge graphs and hinders their wider use** (65).

The semantic unit framework introduced some-instance and every-instance resources as two new types of resources in addition to named-individuals, classes, and properties (50). This allows, besides specifying universal statements and negations without having to use blank nodes and simplifying negations and cardinality statements, modeling them as ABox expressions, also representing questions as statements in the graph (for a discussion, see (10,64)). Instead of named-individuals, one uses some-instance resources of a given class for a given slot of a reference schema to specify that the answer to the question may have any instance of that class at that position in any statement of the type to which the reference schema is associated. If you are asking about universal statements, you would use the every-instance resource instead.

The resulting statement can be stored as a question within the knowledge graph, by classifying its statement resource to instantiate the '*question statement*' class. Since all question statements in a given knowledge graph would reflect the structure of their corresponding reference schema, and since all reference schemata have been created according to the same modeling paradigm, one could develop a **Rosetta Query-Builder** that converts the question into a query that returns a Boolean *true/false* answer if the subject and object-positions contained fully specified inputs (i.e., named-individual resources), or a list of statements as the answer that match the specifications provided by the question statement if one or more of these positions contained underspecified input (i.e., some-instance, every-instance, or class resources).

The same idea provides the basis for developing query interfaces for knowledge graph applications: users can create their own queries with the input forms for creating statements using named-individuals, some-instances, every-instances, classes, and datatypes as inputs for subject and object-positions **without having to know and use any graph query languages**. The Query-Builder translates the question statements into actual graph queries. The generic structure of the reference schemata greatly facilitates the development of procedures for this translation step. By using Boolean operators (AND and OR) and by reusing object resources as subject resources in another question statement, multiple question statements can be combined to form a more complex question.

Question units not only allow you to document questions in the graph, but also to make them operable. This can be used, for example, to document competency questions for a project in a readily actionable way within the knowledge graph itself. The ability to store question statements and make them operable moves the Rosetta Framework another step closer to the concept of **Knowledge Graph Building Blocks** (64) and further enhances its overall **cognitive interoperability**.

## Discussion

Thinking about possible criticisms of the Rosetta Framework, one could mention its underlying idea to consider statements as minimum information units in addition to individual terms and to organize the knowledge graph accordingly. To structure a knowledge graph into statements, each statement can be organized in a nanopublication (66–68) and as a FAIR Digital Object (see also discussion in the context of semantic units (50)). As a consequence, however, the Rosetta Framework requires the specification of a statement class with associated reference schema for each type of statement needed in a knowledge graph. This limitation results from the fact that in the Rosetta Framework, the set of statement classes and associated reference schemata defines the possible proposition-space of the knowledge graph. While this criticism is valid, we want to respond that truly FAIR (meta)data require that every (meta)data statement in a knowledge graph must be FAIR. For a statement to be FAIR, it must be interoperable, and we explained above why this requires the specification of a schema for each statement type. Now, to be truly FAIR, each statement must also reference the schema against which it was modeled for reasons of schematic interoperability, and ideally also which statement type it instantiates. Thus, the need to model each statement type is not unique to the Rosetta Framework, but applies to FAIR knowledge graphs in general.

The difference with other knowledge graph frameworks is that the Rosetta Editor will provide substantial support for creating schemata for new statement types and will not require experience with Semantics, RDF, OWL, or any graph query language, improving cognitive interoperability over other frameworks. In addition, once a reference schema and its corresponding statement class are specified and made available in a schema repository, they can be reused by any other party, reducing the effort required to develop new knowledge graph applications when using the Rosetta Framework. Knowledge graphs that use a dynamic (crowdsourced) approach to knowledge graph construction, and thus do not follow the otherwise typical static information extraction paradigm with a closed set of predefined schemata, particularly benefit from the Rosetta Framework and its Editor, as applying the static approach to dynamic scenarios or domains usually falls short when a new type of statement needs be added to the graph.

Another criticism is that the Rosetta modeling paradigm does not use a logical framework, so you cannot apply reasoning to statements created with it. We agree that it would be desirable to be able to apply reasoning, but we consider it as more important to ensure first the findability and then the general FAIRness and cognitive interoperability of (meta)data. We have therefore chosen the Rosetta modeling paradigm to model statements rather than reality. If reasoning is required, you can always convert (meta)data into a structure that enables reasoning using one of the specified schema crosswalks.

We believe that the Rosetta Framework will help improve the technical and semantic interoperability of (meta)data. Institutions running their knowledge graph applications based on the Rosetta Framework will be able to share their (meta)data and even build a federated virtual knowledge graph with their partners, while data stewardship remains in their own hands, thus ensuring full control over the ethical, privacy, or legal aspects of their data (following Barend Mons' *data visiting as opposed to data sharing* (19)). In addition, by providing the possibility to specify additional access models and to access the (meta)data of the knowledge graph in different formats (including Java and Python data classes), and by making statement classes together with their associated reference schemata and display templates, functions, the Editor, and the Query-Builder openly and freely available, the Rosetta Framework fulfills the recommendations of the EOSC Interoperability Framework for **technical interoperability** (20).

On the other hand, by using Wikidata terms and other controlled vocabularies and ontologies that provide UPRIs and publicly available definitions for their class terms, by tracking and documenting metadata and by providing a way to specify schema crosswalks, by decoupling human-readable data display from machine-actionable data storage, with the former focusing on providing human-readable data views and the latter involving the consistent application of reference schemata that ensure the semantic interoperability of (meta)data, by providing the ability to apply community-specific standards for accessing (meta)data, and by documenting for each statement through its statement instance resource which reference schema had been used to model it, the Rosetta Framework also satisfies the recommendations for **semantic interoperability** (20), while increasing the overall **cognitive interoperability** of its (meta)data and its knowledge graph applications.

The Rosetta Framework further supports cognitive interoperability, by modeling natural language statements rather than reality and by providing easy-to-use tools that remove some layers of complexity and requirements such as having to be proficient in (graph) query languages, since the Rosetta Query-Builder will derive generic CRUD queries based on reference schemata—no need for developers and operators to define schemata and write queries themselves. Moreover, all reference schemata with their crosswalks and all functions, once specified and developed and made available in a repository, can be reused by anyone developing their own knowledge graph applications.

## Related work

[SHACL](#) and [ShEx](#) are shape constraint languages for describing RDF graph structures (i.e., shapes) that identify predicates and their associated cardinalities and datatypes. Shapes can be used for communicating data structures, creating, integrating, or validating graphs, and generating UI forms and code.

[Reasonable Ontology Templates](#) (OTTR) takes up the idea of shapes and uses them as building blocks for knowledge bases. The templates provide an abstraction level that is better suited for managing a knowledge base than the low-level RDF triples or OWL axioms. The templates are stored in a template library which supports reuse and therewith uniform modeling across different knowledge bases. Templates can refer to other templates. With its clear separation of templates and their instantiations in the form of data, OTTR clearly separates between knowledge base design and knowledge base contents. This is comparable to the Rosetta Framework. Tools exist for mapping CSV or relational data to templates.

The Research Data Alliance ([RDA](#)) is developing an InteroperAble Descriptions of Observable Property Terminology ([I_ADOPT](#) (70)), i.e., an Interoperability Framework for representing observable properties that shares some similarities with the Rosetta Framework. The I_ADOPT framework is based on an ontology designed to enhance interoperability between existing models (e.g., ontologies, taxonomies, controlled vocabularies) of semantically describing variables. Variables are understood to be descriptions of something observed or mathematically derived, defined by at least the entity being observed and by corresponding characteristics, but more complex variables can be described as well and require additional entities. One of the difficulties encountered in representing the meaning of such variables is agreeing upon the elements that constitute them. The ontology provides essential atomic components and relations that can be employed to define machine-interpretable FAIR variable descriptions. The I_ADOPT framework does not cover concepts such as units, instruments, methods, and geographic location information, but is confined on the description of the variable itself. It provides templates, i.e., Variable Design Patterns ([VDP](#)s), that are similar to Ontology Design Patterns and provide schemata for specific types of variables.

While there are many similarities between the I_ADOPT and the Rosetta Framework, only the Rosetta Framework models statements instead of a human-independent reality and therefore provides a very generic modeling approach that reflects the structure of natural language statements, thus ensuring the cognitive interoperability of the schemata (=models)—they are easy to apply and easy to understand (e.g., compared to the [step-by-step procedure](#) for minting new variable). In addition, whereas I_ADOPT has a strong focus in environmental research and in encoding measurement and observational (meta)data, the Rosetta Framework is domain-agnostic and can be applied to any type of statement.

We also follow with interest the ongoing [Abstract Wikipedia](#) project and [Wikifunctions](#), which is closely related to it, especially the part relating to **[Constructor Units](#)** and the abstract content language. Constructor Units provide abstract representations of predicate statements and thus follow an idea that shares some similarities with our Rosetta modeling paradigm, with the differentiation of required and optional object-positions. Interesting to us is also the idea to verbalize a constructor unit in more than one sentence for improving its readability and to have several possible sentence-like realizations of it, e.g., in different languages, all being provided in rendering time. Wikifunctions, in turn, is similar to but in dimensions more general than our idea of an open repository for Rosetta Functions.

## Conclusion

Today, with the pressure to rapidly produce new and FAIR (meta)data that meet the requirement to integrate with existing (meta)data sources from different domains, knowledge graph management platforms must support the rapid development of data management solutions that are suitable for everyday research, that demonstrate the added value of knowledge graph technologies, and that can be implemented by smaller projects with limited budgets or even by individual researchers for their personal research knowledge graph (71). This is all the more true given the value of manually authored knowledge from domain experts. According to Bradley's brief history of knowledge engineering (72), the most valuable documentations of (meta)data come from the researchers who created them, and it is the task of knowledge engineers to make it as easy as possible for those researchers to document them as FAIR as possible. We expect that the Rosetta Framework will make this task much easier.

Ideally, a knowledge graph management platform supports the needs of different categories of user groups: **Domain experts**, who are not interested in the internals of the application logic or the underlying data structure of the graph. They are not interested in the additional information that has to be added to make (meta)data machine-actionable, but want to access the information in the graph in intuitive ways with the (meta)data presentation being reduced to the information they are currently interested in.

**Expert users and data scientists**, on the other hand, want to interact with the (meta)data, maybe analyze it using Jupyter Notebooks, R, or other data analysis tools, and require the (meta)data to be easily accessible in the formats they prefer and want the knowledge graph to provide convenient and efficient tools that support their data management operations.

And finally **data stewards, ontology engineers, and application engineers**, who require that the knowledge graph infrastructure can be easily incorporated into the overall software stack of their organizations while retaining full control over their (meta)data's ethical, privacy, or legal aspects (following Barend Mons' *data visiting as opposed to data sharing* (19)). They want the platform to make their lives easier when having to develop new targeted applications on top of the knowledge graph.

With its main idea to clearly distinguish between a generic data storage model, data display models, and data access/export models and to decouple the application data model from displaying data in the UI, the Rosetta Framework supports all this. We argue that this decoupling is essential for being able to solve many of the immanent problems of knowledge management systems. A Rosetta-driven FAIR knowledge graph application, based on the information provided by a set of statement type classes and their associated reference schemata and display templates, allows intuitively making statements about statements, provides users with input-forms and human-readable data views in the UI, enforces graph patterns for internal interoperability and machine-actionability of (meta)data, allows the specification of additional data schemata alongside with corresponding schema crosswalks for access and export with the possibility to add further schemata for newly upcoming standards, and provides CRUD queries derived from the storage models of each statement type that can be intuitively used, combined, and further specified using the Rosetta Query Builder for searching and exploring the graph.

With the low-code Rosetta Editor and an openly accessible reference schema repository, the Rosetta Framework provides a set of resources and tools that increase the overall cognitive interoperability of (meta)data and of knowledge graph applications. Domain experts can define their own Rosetta-driven FAIR knowledge graph applications and developers can access and export their data as JSON, RDF, or CSV.

We think, the time has come that, in addition to focusing on the machine-actionability of (meta)data, we start to focus also on their human-actionability and thus their human-friendliness, creating knowledge graph applications that meet the requirements of cognitive interoperability. The Rosetta Framework is our suggestion for how we could arrive there.

# Acknowledgements

We thank Philip Strömert, Roman Baum, Björn Quast, Peter Grobe, István Míko, and Kheir Eddine for discussing some of the presented ideas. We are solely responsible for all the arguments and statements in this paper. This work was supported by the ERC H2020 Project 'ScienceGraph' (819536). We are also grateful to the taxpayers of Germany.